\documentclass[10pt]{article}
\usepackage[noblocks]{authblk}
\usepackage{tabularx}
\usepackage{graphicx}
\usepackage{multicol}
\usepackage{bm}
\usepackage{units}
\usepackage{amsmath}
\usepackage{amssymb}
\usepackage{color}
\usepackage{ulem}
\usepackage{comment}
\usepackage{lineno}
\usepackage[squaren,Gray]{SIunits}
\usepackage[dvipsnames]{xcolor}
\usepackage[utf8]{inputenc} 
\usepackage{float}
\usepackage{xfrac}
\usepackage{bm}
\usepackage{physics}
\usepackage{ulem}
\usepackage{multirow}
\usepackage{orcidlink}
\newcommand{\mycomment}[1]{}
\usepackage[sorting=none,style=numeric-comp]{biblatex} 
\usepackage{hyperref}
\usepackage{array}
\newcolumntype{L}[1]{>{\raggedright\let\newline\\\arraybackslash\hspace{0pt}}m{#1}}
\newcolumntype{C}[1]{>{\centering\let\newline\\\arraybackslash\hspace{0pt}}m{#1}}
\newcolumntype{R}[1]{>{\raggedleft\let\newline\\\arraybackslash\hspace{0pt}}m{#1}}
\makeatletter
\def\@maketitle{%
  \newpage
  \null
  \vskip 4.em%
  \begin{center}%
  \let \footnote \thanks
    {\huge \bfseries \@title \par}%
    \vskip 0.5em%
    {\bf LiquidO Collaboration}$^{{\scriptscriptstyle (a-z)}}$\thanks{\url{LiquidO-Contact-L@in2p3.fr}}
    {\normalsize 
      \lineskip 0.em%
      \begin{tabular}[t]{c}%
        \@author
      \end{tabular}\par}%
    \vskip 0.em%
\@date%
  \end{center}%
  \par
  \vskip 1.3em}
\makeatother

\newcommand{\Londrina}{Departamento de F\'isica, 
	Universidade Estadual de Londrina, 
	Londrina,
	Brazil}

\newcommand{\PUCR}{Department of Physics, 
	Pontif\'icia Universidade Cat\'olica do Rio de Janeiro, 
	Rio de Janeiro, Brazil}

\newcommand{\Queens}{Department of Physics, 
	Engineering Physics \& Astronomy, 
	Queen's University, Kingston, 
	Canada}

\newcommand{\Prague}{Institute of Particle and Nuclear Physics,
	Charles University,
	Prague, 
	Czech Republic}

\newcommand{\IJCLabSaclay}{Universit\'e Paris-Saclay, 
	CNRS/IN2P3, 
	IJCLab, 
	Orsay, 
	France}

\newcommand{\LPtwoI}{Universit\'e de Bordeaux, 
	CNRS, 
	LP2I Bordeaux, 
	Gradignan, 
	France}

\newcommand{\SUBA}{Nantes Universit\'e, 
	IMT-Atlantique, 
	CNRS, 
	Subatech,
	Nantes, 
	France}
	
\newcommand{\CPPM}{Universit\'e de Aix Marseille, 
	CNRS, 
	CPPM, 
	Marseille, 
	France}

\newcommand{\LNCA}{LNCA Underground Laboratory, 
	CNRS, 
	EDF Chooz Nuclear Reactor, 
	Chooz, France}

\newcommand{\APC}{Universit\'e de Paris Cit\'e, 
	CNRS, 
	APC,
	Paris, France}

\newcommand{\FerraraUni}{Dipartimento di Fisica e Scienze della Terra, 
	Universit\`{a} di Ferrara, 
	Ferrara, 
	Italy}
\newcommand{\FerraraINFN}{INFN, 
	Sezione di Ferrara, 
	Ferrara, 
	Italy}

\newcommand{\Padovauni}{Dipartimento di Fisica e Astronomia, 
	Universit\`{a} di Padova, 
	Padova, Italy}
\newcommand{\Padova}{INFN, 
	Sezione di Padova, 
	Padova, Italy}

\newcommand{\MainzA}{Johannes Gutenberg-Universit\"{a}t Mainz,
	Institut f\"{u}r Physik, 
	Mainz, Germany} 
\newcommand{\MainzB}{Johannes Gutenberg-Universit\"{a}t Mainz,
	Detektorlabor, Exzellenzcluster PRISMA$^+$,
	Mainz, Germany} 	

\newcommand{\MPIK}{Max-Planck-Institut f\"{u}r Kernphysik, 
	Heidelberg, 
	Germany}

\newcommand{\TechnicoIBB}{iBB, 
	Instituto Superior Tecnico, 
	Universidade de Lisboa,
	Lisbon, 
	Portugal}
    
\newcommand{\TechnicoIDMEC}{IDMEC, 
	Instituto Superior Tecnico, 
	Universidade de Lisboa,
	Lisbon, 
	Portugal}

\newcommand{\TechnicoCtwoTN}{CT2N, 
	Instituto Superior Tecnico, 
	Universidade de Lisboa,
	Lisbon, 
	Portugal}

\newcommand{\CIEMAT}{CIEMAT, 
	Centro de Investigaciones Energ\'{e}ticas, Medioambientales y Tecnol\'{o}gicas, 
	Madrid, Spain}

\newcommand{\UniZar}{Centro de Astropart\'{\i}culas y F\'{\i}sica de Altas Energ\'{\i}as (CAPA),
 	Universidad de Zaragoza, 
	Zaragoza, Spain}

\newcommand{\DIPC}{Donostia International Physics Center, 
	Basque Excellence Research Centre, 
	San Sebasti\'an/Donostia, 
	Spain}

\newcommand{\RCNS}{RCNS, 
	Tohoku University, 
	Sendai, Japan}

\newcommand{\Sussex}{Department of Physics and Astronomy, 
	University of Sussex, 
	Brighton, 
	United Kingdom}

\newcommand{\ImpCol}{Department of Chemistry, 
	Imperial College London, 
	London, 
	United Kingdom}
	
\newcommand{\RAL}{Rutherford Appleton Laboratory, 
	Didcot,
	Oxford,
	United Kingdom}

\newcommand{\UCI}{Department of Physics and Astronomy, 
	University of California at Irvine, 
	Irvine, 
	CA, 
	USA}

\newcommand{\UniPennPhys}{Department of Astronomy and Astrophysics, 
	Pennsylvania State University, 
	University Park, 
	PA,
	USA}
\newcommand{\UniPennAstro}{Department of Physics, 
	Pennsylvania State University, 
	University Park, 
	PA,
	USA}

\newcommand{\UniMichigan}{Department of Nuclear Engineering and Radiological Sciences,
 	University of Michigan, 
	Ann Arbor, 
	MI,
	USA}
					
\newcommand{\BNL}{Brookhaven National Laboratory, 
	Upton, 
	NY,
	USA}

\begin{document}
\normalem 

%
%
\title{\bf The Stochastic Light Confinement of LiquidO}
\date{\today}
\mycomment{
}
%
%

%
%
\author[l,z]{J.\,Apilluelo}
\author[b]{L.\,Asquith}
\author[b]{E.\,F.\,Bannister}
\author[k$\alpha$]{N.\,P.\,Barradas}
\author[p]{J.\,L.\,Beney}
\author[k$\beta$]{M.\,Berberan\,e\,Santos}
\author[p]{X.\,de\,la\,Bernardie}
\author[b]{T.\,J.\,C.\,Bezerra\,\orcidlink{0000-0002-0424-7903}}
\author[p]{M.\,Bongrand} 
\author[q]{C.\,Bourgeois}
\author[q]{D.\,Breton}
\author[1]{C.\,Buck\,\orcidlink{0000-0002-5751-5289}}
\author[n]{J.\,Busto}
\author[q]{K.\,Burns}
\author[q,c,2]{A.\,Cabrera\,\orcidlink{0000-0001-5713-3347}}
\author[p]{A.\,Cadiou}
\author[l]{E.\,Calvo}
\author[f]{E.\,Chauveau}
\author[b]{B.\,J.\,Cattermole}
\author[h]{M.\,Chen}
\author[i]{P.\,Chimenti} 
\author[x$\alpha$,x$\beta$]{D.\,F.\,Cowen}
\author[r$\alpha$]{S.\,Dusini\,\orcidlink{0000-0002-1128-0664}} 
\author[b]{A.\,Earle}
\author[k$\alpha$]{M.\,Felizardo}
\author[i]{C.\,Frigerio\,Martins}
\author[z]{J.\,Gal\'an}
\author[z]{J.\,A.\,Garc\'ia}
\author[q]{R.\,Gazzini}
\author[b]{A.\,Gibson-Foster}
\author[m$\alpha$]{C.\,Girard-Carillo}
\author[1]{B.\,Gramlich}
\author[2,r$\beta$]{M.\,Grassi}
\author[b]{W.\,C.\,Griffith}
\author[u]{J.\,J.\,G\'omez-Cadenas}
\author[p]{M.\,Guiti\`ere}
\author[p]{F.\,Haddad}
\author[b]{J.\,Hartnell} 
\author[d]{A.\,Holin}
\author[z]{I.\,G.\,Irastorza}
\author[a]{I.\,Jovanovic\,\orcidlink{0000-0003-0573-3150}}
\author[k$\alpha$]{A.\,Kling}
\author[m$\alpha$]{L.\,Koch\,\orcidlink{0000-0002-2966-7461}}
\author[b]{P.\,Lasorak}
\author[q,c]{J.\,F.\,Le\,Du}
\author[h]{C.\,Lefebvre}
\author[p]{F.\,Lefevre}
\author[q]{P.\,Loaiza}
\author[b]{J.\,A.\,Lock}
\author[z]{G.\,Luz\'on}
\author[q]{J.\,Maalmi}
\author[j]{J.\,P.\,Malhado}
\author[e$\alpha$,e$\beta$]{F.\,Mantovani}
\author[k$\alpha$]{J.\,G.\,Marques}
\author[f]{C.\,Marquet} 
\author[z]{M.\,Mart\'inez}
\author[q,l]{D.\,Navas-Nicol\'as\,\orcidlink{0000-0002-2245-4404}}
\author[t]{H.\,Nunokawa} 
\author[2]{M.\,Obolensky} 
\author[g]{J.\,P.\,Ochoa-Ricoux\,\orcidlink{0000-0001-7376-5555}} 
\author[k$\beta$]{T.\,Palmeira}
\author[l]{C.\,Palomares} 
\author[k$\beta$]{B.\,Pedras}
\author[d]{D.\,Petyt}
\author[p]{P.\,Pillot}
\author[f]{A.\,Pin}
\author[b]{J.\,C.\,C.\,Porter} 
\author[f]{M.\,S.\,Pravikoff\,\orcidlink{0000-0002-7088-4126}}
\author[k$\beta$]{N.\,Rodrigues}
\author[f]{M.\,Roche}
\author[y]{R.\,Rosero}
\author[s]{B.\,Roskovec}
\author[q]{N.\,Roy}
\author[z]{M.\,L.\,Sarsa}
\author[m$\beta$,1]{S.\,Schoppmann\,\orcidlink{0000-0002-7208-0578}}
\author[r$\alpha$,r$\beta$]{A.\,Serafini}
\author[d]{C.\,Shepherd-Themistocleous}
\author[b]{W.\,Shorrock\,\orcidlink{0000-0002-7221-1910}}
\author[k$\gamma$]{M.\,Silva}
\author[q]{L.\,Simard}
\author[u]{S.\,R.\,Soleti}
\author[m$\alpha$,m$\beta$]{H.\,Th.\,J.\,Steiger}
\author[p]{D.\,Stocco}
\author[e$\alpha$,e$\beta$]{V.\,Strati}
\author[p]{J.\,S.\,Stutzmann}
\author[v]{F.\,Suekane}
\author[m$\alpha$]{A.\,Tunc}
\author[b]{N.\,Tuccori\,\orcidlink{0000-0002-2868-5887}}
\author[l]{A.\,Verdugo}
\author[p]{B.\,Viaud}
\author[m$\alpha$]{S.\,M.\,Wakely\,\orcidlink{0000-0002-2919-8159}}
\author[m$\alpha$]{A.\,Weber\,\orcidlink{0000-0002-8222-6681}}
\author[x$\beta$]{G.\,Wendel}
\author[a]{A.\,S.\,Wilhelm\,\orcidlink{0000-0002-0664-0477}}
\author[y]{M.\,Yeh}
\author[p]{F.\,Yermia}
%
%
%
%
\affil[a]{\UniMichigan} 
\affil[b]{\Sussex} 
\affil[c]{\LNCA} 
\affil[d]{\RAL} 
\affil[e$\alpha$]{\FerraraINFN} 
\affil[e$\beta$]{\FerraraUni} 
\affil[f]{\LPtwoI} 
\affil[g]{\UCI} 
\affil[h]{\Queens} 
\affil[i]{\Londrina} 
\affil[j]{\ImpCol} 
\affil[k$\alpha$]{\TechnicoCtwoTN} 
\affil[k$\beta$]{\TechnicoIBB} 
\affil[k$\gamma$]{\TechnicoIDMEC} 
\affil[l]{\CIEMAT} 
\affil[m$\alpha$]{\MainzA} 
\affil[m$\beta$]{\MainzB} 
\affil[n]{\CPPM} 
\affil[p]{\SUBA} 
\affil[q]{\IJCLabSaclay} 
\affil[r$\alpha$]{\Padova} 
\affil[r$\beta$]{\Padovauni} 
\affil[s]{\Prague} 
\affil[t]{\PUCR} 
\affil[u]{\DIPC} 
\affil[v]{\RCNS} 
\affil[x$\alpha$]{\UniPennPhys} 
\affil[x$\beta$]{\UniPennAstro} 
\affil[y]{\BNL} 
\affil[z]{\UniZar} 
%
%
\affil[$\phantom{0}$]{---------------} 
\affil[1]{\MPIK} 
\affil[2]{\APC} 
%
%
\maketitle

\noindent
\begin{center}
	\parbox{0.95\textwidth}
	{\bf \small \noindent
    Light-based detectors have been widely used in fundamental research and industry since their inception in the 1930s. 
    The energy particles deposit in these detectors is converted to optical signals via the Cherenkov and scintillation mechanisms that are then propagated through transparent media to photosensors placed typically on the detector's periphery, sometimes up to tens of metres away.
    LiquidO is a new technique pioneering the use of opaque media to stochastically confine light around each energy deposition while collecting it with an array of fibres that thread the medium. 
    This approach preserves topological event information otherwise lost in the conventional approach, enabling real-time imaging down to the MeV scale. 
    Our article demonstrates LiquidO's imaging principle with a ten-litre prototype, revealing successful light confinement of 90\% of the detected light within a 5\,cm radius sphere, using a custom opaque scintillator with a scattering length on the order of a few millimetres.
    These high-resolution imaging capabilities unlock opportunities in fundamental physics research and applications beyond.
    The absolute amount of light detected is also studied, including possible data-driven extrapolations to LiquidO-based detectors beyond prototyping limitations. Additionally, 
    LiquidO's timing capabilities are explored through its ability to distinguish Cherenkov light from a slow scintillator.
}
\vspace{0.5cm}
\end{center}
\begin{multicols*}{2}

\section{Introduction}\label{sec:intro}

Since the 1930s, 
light-based detectors have played a central role in a wide range of physics disciplines spanning fundamental particle physics, astronomy, nuclear physics, and recent developments in medical imaging~\cite{PDG,Shwartz:2017efz,Ronda_2016}.
In these detectors, the energy deposited by particles is converted into light via the scintillation~\cite{Birks_1964} and Cherenkov~\cite{Cherenkov_1934} mechanisms. 
This light is then propagated through the detector's medium to single-photon sensors, typically located on the periphery.
 


Light-based detectors have typically been optimised to rely primarily on either Cherenkov or scintillation light. The light yield of organic liquid scintillators is around 10,000 photons per MeV, making these detectors well-suited for detecting low-energy particles in the MeV range. However, the abundant and isotropic light, combined with the nanosecond-scale response of the large photosensors typically used to detect it, severely limits the ability to resolve topological information. Detectors relying primarily on scintillation light also tend to have limited vertex resolution, $\mathcal{O}(10\,\text{cm})$, and generally offer little to no directionality discrimination. 
Some of these limitations can be mitigated through segmentation, which grants some amount of topological information in a pixelated fashion. 
This, however, comes at the expense of drawbacks such as light losses, increased cost, and potentially higher radioactive background levels.
Moreover, segmenting finely enough to resolve the complete topological information of interactions in the few-MeV range remains 
challenging.

Relying on the prompt Cherenkov light may preserve some degree of topological information, as it is emitted in a conical shape centred along the trajectory of a charged particle above the kinematic threshold~\cite{Cherenkov_1934,Frank_Tamm_1937}.
However, the light yield is approximately a factor of 100 lower than that of an optimised liquid scintillator, which is consequently preferred for achieving low energy thresholds. 
Moreover, while the topological information can be exploited in track-like energy depositions, the pattern is expected to wash out in particles undergoing strong random scattering~\cite{Cherenkov4Electrons}.


This article presents a new approach for light-based detectors called LiquidO~\cite{LiquidO_Website} and demonstrates its performance with a small prototype. LiquidO breaks with the conventional paradigm of transparency by using an opaque scintillator, resulting in the {\it stochastic confinement} of optical photons near their creation point.
This approach, conceived in 2012-2013, was first 
reported~\cite{CERN-Seminar2019@Zenodo,LiquidO_2019} in 2019 upon accomplishing a first proof-of-principle demonstration.

The publication's contents are organised as follows. 
Section~\ref{sec:liquid} outlines the LiquidO detection technique, emphasising its high-resolution and particle discrimination capabilities for 
physics applications at different energies.
Section~\ref{subsec:setup} describes the experimental setup used to derive the results presented subsequently, while Section~\ref{subsec:analysis} outlines the methods used to analyse the data. Section~\ref{subsec:confinement} presents a data-driven demonstration of stochastic light confinement, which is LiquidO's defining feature.
Section~\ref{subsec:lightamount} reports the absolute amount of light measured in the prototype and derives data-driven scaling prospects for other LiquidO-like detectors.
Finally, Section~\ref{subsec:timing} discusses the time response, including the ability to identify Cherenkov and scintillation light, which is expected to provide additional discriminatory power. 
The Appendices contain complementary detailed information on 
the event selection (\ref{annexe:preselection}), 
waveform reconstruction (\ref{annexe:recozor}), 
detector simulation (\ref{annexe:simulation}),
detector response uniformity (\ref{annexe:efficiencycorrection}),
and a simplified analytical scattering model (\ref{annexe:analyticalmodel}). 
Both the simulation and the model were developed to gain insight into the detector's response, light yield and patterns resulting from stochastic light confinement. 

\section{LiquidO's Detection Principle}
\label{sec:liquid}

A LiquidO detector consists of a volume filled with an opaque medium where, regardless of their production mechanism, photons undertake random-walk trajectories and thus form a {\it light ball} around the location of each energy deposition. 
Since light cannot travel far from its origin, the detector must be endowed with a dedicated system to collect it inside the medium. Today's technology enables this by utilising a dense array of wavelength-shifting or scintillating fibres that traverse the medium.
The direction in which fibres are oriented is referred to as the readout axis.
Up to three orthogonal readout axes may be considered, but typically one may suffice. 
The fibres collect the emitted photons and trap them following wavelength-shifting, channelling them for detection to a readout system relying on fast photosensors, typically semiconductor-based silicon photomultipliers (SiPMs), located at the end(s) of each fibre. Depending on the physics goal, the readout specifics may vary, whereas the exploitation of stochastic light confinement remains LiquidO's defining feature.

The exact topology of the light ball results 
from the competition between scattering, detection, and losses.
The latter two can be regarded as the absorption by the fibres and the medium, respectively.
Hence, maximising light collection by mitigating absorption losses necessitates the most efficient possible photon detection, driven by a balance between the mean scattering length and the density of fibres used for a given absorption length.

The ideal opaque scintillator in a LiquidO detector has an absorption length of several metres~\cite{Birks_1964} but a scattering length up to four orders of magnitude smaller, down to the millimetre scale~\cite{LiquidO_2019}.
LiquidO aims to exploit dominant elastic scattering mechanisms, such as Rayleigh and Mie scattering, while minimising absorption.
This has been achieved by modifying transparent scintillators~\cite{Cabrera_DRD2,wagner_2018,Buck_2019,Schoppmann_2023,Schoppmann_2024}, although other approaches remain possible and are under exploration.

An electron with energy around 1\,MeV in a conventional liquid scintillator (density $\sim$0.9\,g/cm$^3$) deposits its energy over a short ionisation track (length $\le$5\,mm) culminating in a Bragg peak. 
In an opaque medium with a comparable scattering length, the result is effectively a single light ball.
In contrast, a $\gamma$-ray manifests as a chain of light balls, each produced by a low-energy electron-recoil arising from each Compton scatter with a final photo-electric conversion, also producing an electron-recoil.
Similarly, a positron (e$^+$) is the combination of an electron-like central light ball resulting from the loss of kinetic energy followed by two light-ball chains corresponding to the two 511\,keV annihilation $\gamma$-rays emitted back-to-back. 
Ref.~\cite{LiquidO_2019} shows a few examples of these events, which result in distinct topological signatures of single (electron) or multiple ($\gamma$-ray or positron) light balls arranged differently in space. 
Consequently, LiquidO unlocks the ability to discriminate between electrons, $\gamma$-rays, and positrons down to the low MeV range on an event-by-event basis and with high efficiency for the first time.
This topological characterisation is achieved without physical segmentation of the detector and the associated detrimental effects of introducing inactive material. 
Instead, the detector becomes effectively self-segmenting thanks to the medium's opacity.
These imaging capabilities all hinge on detecting the primitive response topology of point-like energy depositions -- a single light ball -- motivating the experimental methodology described in the next section. 

As energies increase to tens of MeV, charged particles produce a continuous sequence of light balls that look increasingly track-like~\cite{LiquidOIEEE24}. 
When showering occurs, which becomes dominant with electrons above a hundred MeV due to bremsstrahlung, LiquidO can provide high-definition images~\cite{LiquidO_GeV} that rival the best detectors in the field.
Additionally, LiquidO is expected to yield a 
vertex resolution of $\mathcal{O}(1\,\text{mm})$ compared to $\mathcal{O}(10\,\text{cm})$ in conventional light-based detectors, as recently corroborated by another LiquidO prototype~\cite{LiquidO-LIME, Wilhelm_2024}. 

LiquidO's approach has other notable advantages.
First, the nascent field of opaque scintillation, pioneered originally for LiquidO~\cite{Cabrera_DRD2}, presents a tantalising landscape of new materials and chemical formulations to explore, which could offer unprecedented opportunities beyond today's capabilities.
Second, 
the reduced light path in the scintillator medium leads to higher resilience against absorption effects, leading to increased light collection and better adaptability to heavy elemental medium loading~\cite{Buck_2016}.
This extends the range of possible applications of LiquidO technology~\cite{LiquidO_2019} to situations where doping with new materials or at significantly higher levels than what is currently achievable is desirable.
Third, unlike most detectors based on PMTs or electron-drift, LiquidO's SiPM-based photodetection enables the possibility of a strong magnetic field, which can be exploited for purposes such as charge tagging~\cite{LiquidO_GeV}.
Fourth, LiquidO is also expected to benefit from the traditional scintillators' pulse shape discrimination~\cite{Birks_1964} (PSD), as electron-like protons and alpha particles lead to different deexcitation time profiles. 
As discussed later, LiquidO's time resolution is expected to comfortably meet today's requirements for plausible PSD.

Overall, the foreseen capabilities of this type of detector make it a compelling option in several areas of particle physics and related fields.
It is currently being studied for several next-generation experiments in neutrino physics~\cite{LiquidO_GeV,LiquidOGeoNu,AM-OTech&CLOUD,SuperChooz}.
There are also ongoing explorations in 
reactor monitoring~\cite{WebAMOTech},
medical applications~\cite{LPET},
high-energy collider-based calorimetry~\cite{PowderO}, 
$\beta\beta$ nuclear decay~\cite{
LiquidO-DBD-FranceI,
LiquidO-DBD-FranceII,
LiquidO-DBD-Paper,
NuDoubt}, and
astrophysical explorations~\cite{Soleti_2025}.


\begin{figure*}[ht]
    \centering
    \includegraphics[width=0.9\linewidth]{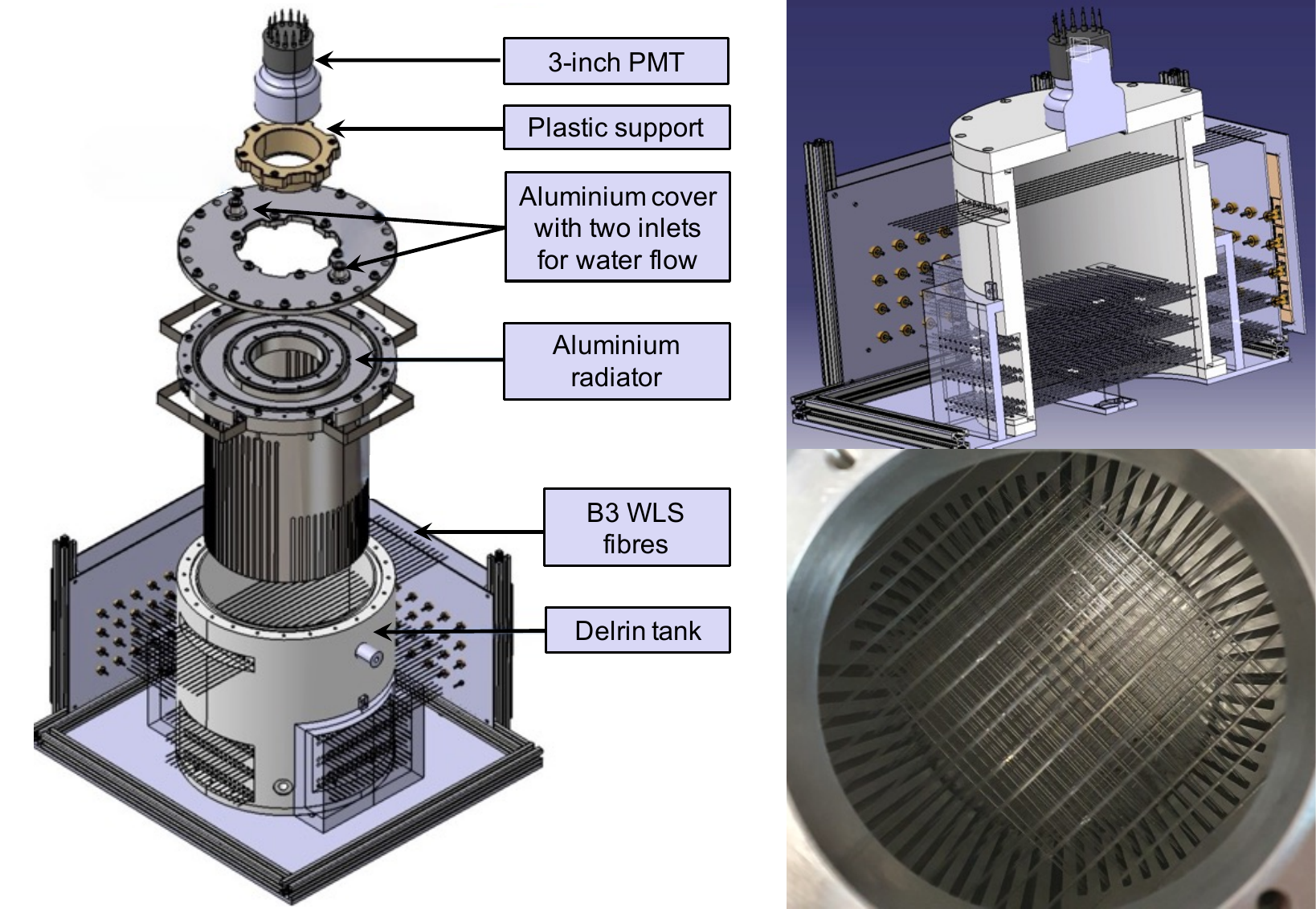}
    \caption{\small 
    {\bf The Mini-LiquidO Detector Prototype System.}
    The detector consists of a cylindrical volume of approximately 10 litres (25 cm diameter, 24 cm height).
    A tuneable monoenergetic single-electron beam enters the detector through a thin aluminised reflective (92\%) 24\,$\mu$m Mylar layer in the middle of the bottom lid.  
    An internal cylindrical aluminium radiator (water-based closed circuit) controls the temperature while permitting the liquid to circulate freely. 
    Light is collected using 208 optical double-cladding wavelength-shifting fibres (Kuraray B-3) running along two mutually orthogonal directions arranged in 6 rows of 8 fibre groups with 4 fibres per group.
    Only one fibre end per group (1:8) is instrumented.
    The average distance between fibre groups is 15\,mm and the distance between fibres is 7.5\,mm.
    An additional row of 16 fibres (2 fibres per group) is placed at the top to monitor far-reaching photons up to 20\,cm away from the beam spot.
    Again, only one fibre end per group is instrumented.
    Close to the upper layer of fibres, there is a 3-inch PMT (HZC-Photonics XP72B20) embedded in the top lid as a reference photodetector.
    The separation (non-instrumented gap) between the upper row and the one below is 10\,cm.
    The left panel shows an exploded view of the setup with all the components. 
    The top right panel shows a cross-section of the detector.
    The bottom right panel is a photo of the inside of the detector when the top PMT is removed, showing the grid of fibres and the reflective Mylar underneath. 
    The full prototype, including the readout cards, is placed inside a dark box to minimise parasitic ambient light.
    }
    \label{fig:setup}
\end{figure*}


\begin{table*}[ht]
    \centering
    \begin{tabular}{|l|c|c|c|c|c|c|}
    \hline
    {\bf Name}  
        & {\bf Operation} 
        & {\bf Solvent} 
        & {\bf Fluor} 
        & {\bf Loading} 
        & {\bf Absorption} 
        & {\bf Scattering} \\
    & {\bf Temperature} & & {\bf (g/L)} & {\bf(weight-\%)} & {\bf Lengths (m)} & {\bf Length (m)} \\
    \hline
    Water   & 20\,\textcelsius & de-ionised water   & -- & --   & $>$10 / $>$10 & $>10$ \\
    LAB     & 20\,\textcelsius & LAB                & -- & --   & $>$0.5 / $>$10 & $>10$ \\
    LAB+PPO & 20\,\textcelsius & LAB                & 3 (PPO)   & -- & $>$0.5 / $>$10 & $>10$ \\
    NoWaSH-Transparent   & 40\,\textcelsius  & LAB   & 3 (PPO)   & 20 (wax) & $>$0.5 / $>$10 & $>1$ \\
    NoWaSH-Opaque        & 5\,\textcelsius   & LAB   & 3 (PPO)   & 20 (wax) & $>$0.5 / $>$10 & $\sim$0.001 \\
    \hline
    \end{tabular}
    \caption{\small 
    {\bf Detection Media used in Mini-LiquidO.}
    LiquidO, like other liquid-based detectors, benefits from the versatility of filling the same detection volume with different liquid media with distinct properties in response to radiation.
    Mini-LiquidO exploits this feature to enable insightful relative response comparisons in the setup (same readout, geometry, etc.) when exposed to the same MeV electrons.
    Thus, the detector was filled with pure water and LAB-based ({\it linear-alkyl-benzene}) liquid scintillators.
    Since a few properties of the scintillator may be temperature dependent, such as its light yield of (roughly -0.4\% per \textcelsius), the detector's radiator controls and stabilises the volume's temperature to $\sim$0.1\textcelsius.
    Each element of the liquid scintillator formulation is added and characterised to ensure a gradual and complete understanding of the detector's optical model and the scintillators.
    LAB alone (not an optimised scintillator: low and slow light yield) is used before and after adding the main PPO ({\it 2,5-diphenyloxazole}) fluor.
    LAB+PPO is doped with 20\% of paraffin wax to achieve NoWaSH~\cite{Buck_2019} (i.e. New opaque Wax Scintillator Heidelberg).
    Lowering the temperature causes the doped wax to crystallise so that the NoWaSH can go from somewhat transparent and liquid ($\ge$35\,\textcelsius; operated at 40\,\textcelsius) to opaque and solidified ($\le$15\,\textcelsius; operated at 5\,\textcelsius).
    The control of the temperature enables characterisation during the phase transition while its optical absorption does not change sizeably, as determined experimentally.  
    In contrast, the scattering length, dominated by Mie scattering, changes dramatically as optical photons are predominantly elastically scattered off the paraffin crystal structure.
    The average optical properties of the materials at 370\,nm and 430\,nm are reported, where the former is most representative in our detection setup.
    For the scattering length, the values are approximately identical hence only one value is provided~\cite{yeh_2024_12745269}.
    Our setup used non-purified LAB, resulting in shorter absorption lengths. 
    Purified LAB can yield longer absorption lengths, up to order 5\,m at 370\,nm.
    }
     \label{tab:liquids}
\end{table*}

\section{Results}\label{sec:results}

\subsection{Experimental Setup \& Data Acquisition}\label{subsec:setup}

The experimental demonstration presented next is obtained using the Mini-LiquidO prototype detector, a cylindrical delrin vessel of approximately 10 litres volume, shown in \autoref{fig:setup}.
Mini-LiquidO is exposed to a high-energy resolution spectrometer, initially developed for SuperNEMO~\cite{Ref_BeamSource}, providing single electrons with millimetre-scale positional precision. 
The beam energy can be varied between 0.4 and 1.8\,MeV by selecting electrons from a ${}^{90}\text{Sr}$ source in an adjustable magnetic field. 
The beam impinges vertically from the bottom, coinciding with the cylindrical axis of the vessel.
The lower region of the detector is called the interaction region, where most instrumented fibres are located, as explained below. 

The bottom lid of the detector has a small cut-out at its centre that is covered by a 24 $\mu$m thin aluminised Mylar membrane. 
The beam enters the detector through this membrane, losing only a very small amount of energy ($\le$7\,keV). 
As a result, the electrons deposit most of their energy within millimetres of their entry point, referred to as the beam spot, producing point-like energy depositions at the bottom of the detector. 
The side of the Mylar membrane facing the inside is aluminised and has an average reflectivity of $(92\pm1)$\% across the relevant wavelength region.

The data taken with the detector included runs with two transparent media and one opaque medium, strategically chosen to enable meaningful relative response comparisons.
As listed in \autoref{tab:liquids}, the transparent media are water (CAS no.:~7732-18-5) and {\it linear-alkyl-benzene} (LAB, CAS no.:~67774-74-7) solvent with and without the addition of {\it 2,5-diphenyloxazole} (PPO, CAS no.:~92-71-7) as fluor or primary wavelength shifter, at a nominal concentration of 3\,g/l.
The opaque medium is a scintillator made of LAB+PPO at the same concentration mixed with paraffin wax (CAS no.:~8002-74-2).
This formulation is referred to as NoWaSH~\cite{Buck_2019}.
NoWaSH at 20\,\% wax loading by weight, labelled as NoWaSH-20, was selected thanks to its advantageous behaviour for LiquidO prototyping.
At warm temperatures ($\ge$35\,\textcelsius; operated at 40\,\textcelsius), it is effectively transparent (referred to as NoWaSH-Transparent), while at lower temperatures ($\le$15\,\textcelsius; operated at 5\,\textcelsius) it is opaque (referred to as NoWaSH-Opaque).
This behaviour is caused by the crystallisation of the paraffin into a homogeneous suspension, which also results in some degree of solidification.

The medium's temperature is measured and controlled by an external chiller system that circulates water in an internal metallic radiator surrounding the central volume.
A redundant probe submerged in the medium provides an independent reading of its temperature.
Upon equilibrium, both temperature measurements show consistent values.

The light inside Mini-LiquidO is collected using two complementary systems: 
a 3-inch HZC-Photonics (XP72B20)~\cite{XP72B20} photomultiplier at the top furthest position from the interaction region ($\sim$23\,cm away) 
and 
a grid of 208 Kuraray B-3~\cite{Kuraray} wavelength-shifting fibres covering mainly the interaction volume.
The photomultiplier readout emulates the conventional approach relying on transparency, while the fibre grid corresponds to LiquidO.
Simultaneous direct relative response comparison between both systems is exploited.
This capability is enhanced by placing 16 fibres very close to the face of the photomultiplier to ensure maximal accuracy.

The 208 fibres traversing the vessel are arranged in two orthogonal directions parallel to the floor's plane, covering the interaction region's main volume, as shown in \autoref{fig:setup}.
The average distance between fibres is 7.5\,mm.
Considering the symmetry in the light collection, only 56 out of the 208 fibres are read out; hence, the average detected light is one out of four (1:4), or 25\% of what is collected by the full array. 
This approach enables 75\,\% cost-savings in the readout. 
Moreover, only one side of the instrumented fibres is read out by a Hamamatsu S13360-1350PE SiPM~\cite{SiPM} using a 
custom-made 
light collector (air-coupled),
while the other side is not instrumented and is left cut open to avoid any possible reflections.
Accordingly, the integral average light detection to collection ratio is reduced by an additional factor of 2 for a combined eight-fold (1:8) reduction overall.  
It should be noted that the non-instrumented fibres still play a role in the light detection process, as photons are still absorbed by the fibres regardless of whether they are read out. This is important for the response scaling analysis addressed in Section~\ref{subsec:lightamount}. 
Overall, this readout strategy reduces the light level per channel, ensuring accurate control of possible response
systematics. 


Custom front-end electronics have been developed 
for the detector to ensure a high level of performance.
Individual channel boards (called ``SiC'' or the SiPM Card) house custom light collectors, which hold the end of the fibre, the SiPM, and a Monolithic Microwave Integrated Circuit Radio Frequency amplifier ($\sim$20\,dB) that also optimises the signal shape for digitisation.
Up to 32 SiCs 
are mounted on a motherboard (USB-driven; called ``SiBB'' or the SiPM Battery Board), which provides all the necessary services, including the temperature-regulated voltage supply for the SiPMs.
The SiC amplifier drives the 3-metre coaxial signal cables from the SiBB to the 64-channel waveform digitiser (WaveCatcher system: 12\,bits dynamic range and 3.2\,GHz sampling)~\cite{Breton7097545,wavecatcher}. 
The overall single-photon time resolution is expected to be 
$\mathcal{O}(100\,\text{ps})$~\cite{PUILL2012354}.

\subsection{Data Analysis}\label{subsec:analysis}
\begin{figure*}[t]
    \centering
    \includegraphics[width=0.49\linewidth]
    {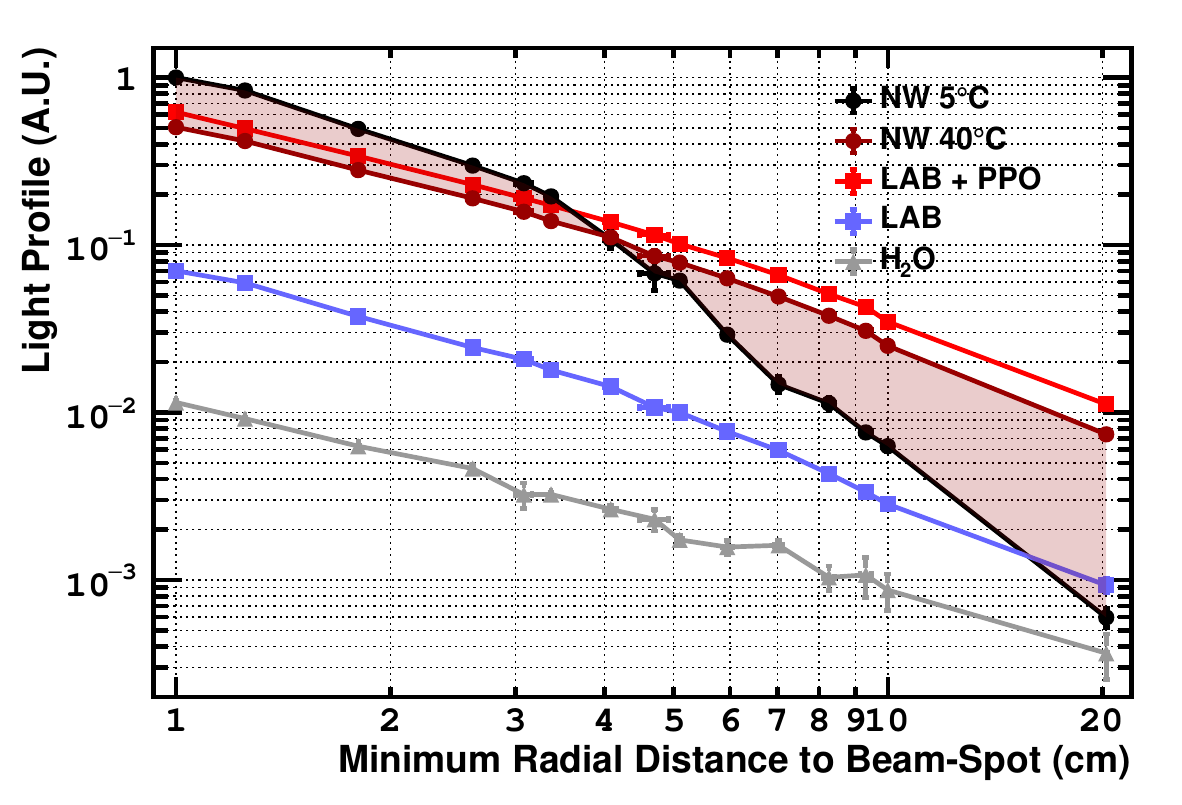}
    \includegraphics[width=0.49\linewidth] 
    {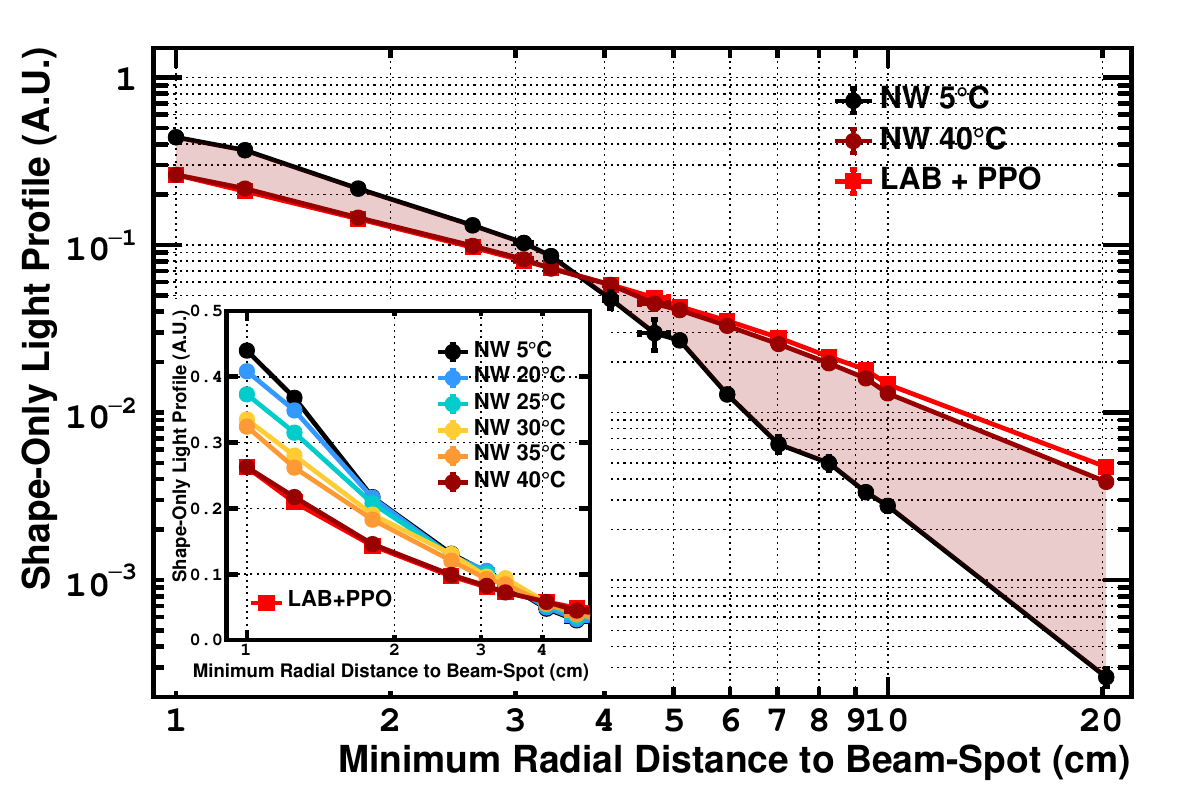}
    \caption{\small 
    {\bf Measured Light Profiles for Different Media.}
    The plots show the light detected as a function of the minimal distance between each fibre and the beam spot after LN subtraction.
    The relative uniformity efficiency correction, detailed in Appendix~\ref{annexe:efficiencycorrection}, is applied in all cases.
    Fibres with similar minimal distances are clustered into single data points.
    The left panel shows the response light profiles in arbitrary units for 
    Cherenkov only in water (grey),
    LAB (blue),
    LAB+PPO (red)
    and NoWaSH (NW) in its transparent and opaque configurations.
    The response integral varies dramatically between the different configurations, as expected.
    There is a slight increase in response for the LAB+PPO case relative to NoWaSH-Transparent due to the wax content, as detailed in the text. 
    Despite some expected forwardness, the Cherenkov light profile in water resembles that of the LAB+PPO, suggesting a high degree of electron random scattering~\cite{Cherenkov4Electrons}.
    The right panel exhibits some of the same curves as the left panel after normalising their area to unity (i.e. shape-only).
    The LAB+PPO provides a control sample that is perfectly transparent and has isotropic emission. 
    The red-shaded areas highlight the difference between the NoWaSH-Transparent and NoWaSH-Opaque curves, showing an enhancement of light collection within the first few centimetres and a severe reduction of light available for detection further away in the opaque case. 
    After only 8\,cm, the relative response falls more than two orders of magnitude.
    The inset (right) shows a NoWaSH-Opaque temperature scan, illustrating the gradual formation of LiquidO's light ball.
    Each profile includes \textgreater50,000 events at 1.8\,MeV beam energy.
    }
    \label{fig:lyorbitals}
\end{figure*}
The results shown throughout this publication are obtained exclusively using electron beam events.
The associated energy depositions occur at well-known time windows during which the readout is externally triggered by tagging the impinging single-electron with a thin plastic scintillator. 
False triggers are caused primarily by trigger electronics' spurious noise, while cosmic-ray events occurring during a beam trigger window constitute one of the main physics backgrounds.
Most of these events are tagged and successfully rejected, as 
detailed in Appendix~\ref{annexe:preselection}. 
This allows the collection of high-purity electron samples that can be statistically combined for response stability and topology studies. 

Upon each trigger, all readout channels are digitised, including the 56 SiPM-equipped channels, the 2 trigger-photomultipliers, and the top PMT, leading to a 320\,ns long waveform per channel, sampled at 3.2 GHz.
The waveform digitiser data enable light pulses to be identified and reconstructed (time, charge and number of detected photons) using a de-convolution method developed originally for the Double Chooz experiment~\cite{DoubleChooz_2019}, where each pulse is fitted by a combination of single-photon hits~\cite{Abe_2014, Hourlier_2016, Pronost_2015}. 
This method, described in Appendix~\ref{annexe:recozor}, performs exceptionally well at identifying each pulse at low photon count rates. 
As a result of the large statistics accumulated, the response uncertainties are dominated by systematic effects caused by small background contamination ($\leq$5\,\%) and some observed variations ($\leq$5\,\%) in the reproducibility of the setup.
Together, these systematic uncertainties were conservatively estimated to be 10\,\%.

After the waveform reconstruction, selected beam events contain a fraction of light that is uncorrelated to the energy deposited by the electron, caused predominantly by the SiPM's dark noise but also by natural radioactivity or residual light leaks.
Given the high sensitivity of detected light envisaged in our prototype, it is crucial to understand and remove this ``light noise'' (LN) contamination to characterise its performance, especially in terms of absolute light response.
The LN contamination is constant in time and independent of the physics events, allowing for statistical subtraction.
The average LN level in each channel is measured to be around $\sim 3\cdot 10^{5}$ single-photons per second. 
This quantity shows a clear positive correlation with the temperature of the SiPMs ($\sim$\,30\,\textcelsius), as expected for dark noise, implying that a sizeable reduction in LN could be achieved by cooling.
For the transparent scenario, the LN rate is more acute in the rows of fibres close to the photomultiplier. 
This is consistent with the higher radioactivity in its glass and photocathode, as well as by a small amount of ambient light entering through the edges of the detector lid. 
Once the medium is opaque, the average LN is reduced, demonstrating the expected self-shielding of LiquidO against external light~\cite{LiquidO_2019}.

The results in this publication incorporate a robust LN subtraction method that accounts for even slight variations on short timescales. 
Channel-wise LN levels are determined {\it in situ} for each run, typically lasting for several hours, as the average light seen in the first 10-50\,ns of each readout window, where no physics signals are present.
The light measured in this off-time signal window is subtracted from the number of photoelectrons (PE) measured in the in-time signal window, typically 50-180\,ns.
As expected, the number of hits before and after this in-time signal window is consistent with zero after LN subtraction. 
The statistical subtraction approach is validated with random triggers with no physics injected in the in-time signal window. 
Consistent results are obtained when the LN is determined on an event-by-event basis. 
However, the lower statistics and the proven high stability of the system make the run-by-run determination more desirable.
The statistical uncertainty of the measured LN is considered in the analysis. 

The results are obtained by making relative comparisons between the data, using the transparent media as a reference to explore LiquidO's exploitation of the opaque medium.
This approach allows for robust result extraction with minimal dependency on detector simulations thanks to a cancellation of systematic effects common to all samples, namely those related to the beam, the trigger and the readout (fibres and SiPMs).  
Still, a simulation provides insight and quantitatively corroborates certain observations and response-related effects, as detailed in Appendix~\ref{annexe:simulation}.

\begin{figure*}[!ht]
    \centering
    \includegraphics[width=0.8\linewidth]{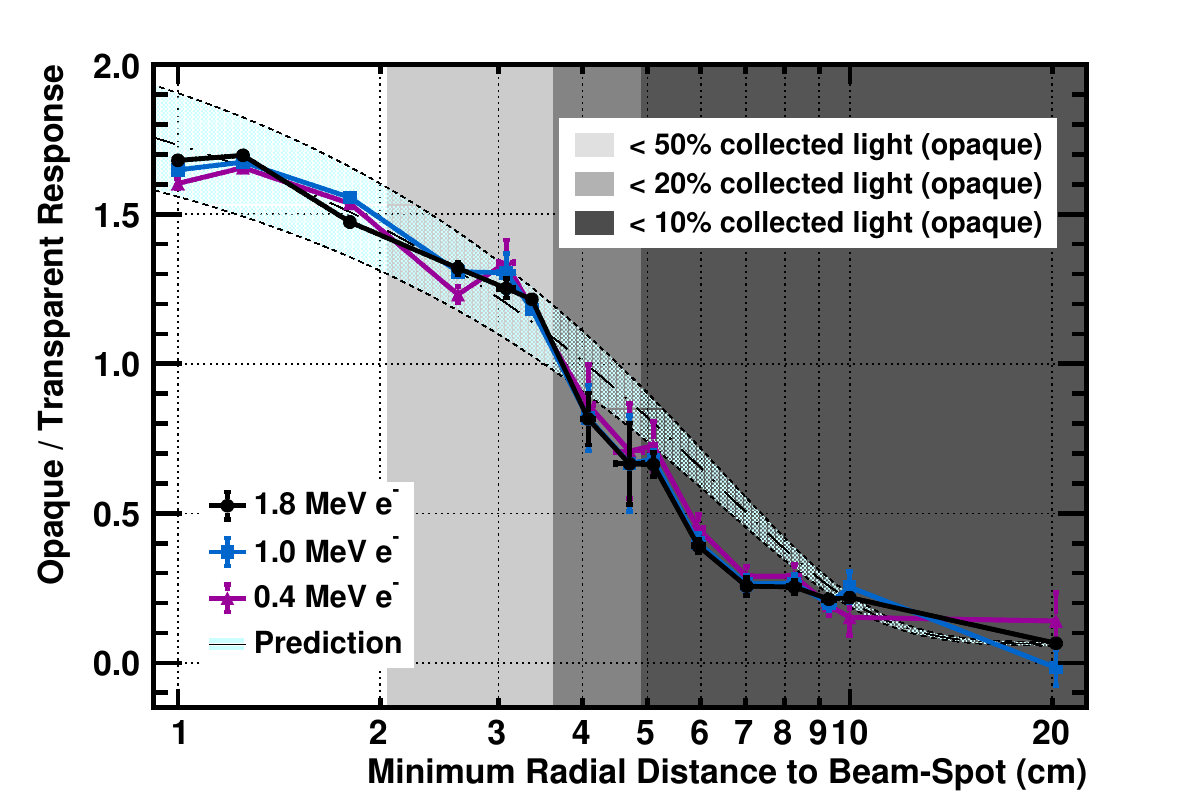}
    \caption{\small 
    {\bf LiquidO's Stochastic Light Confinement Shape.}
    The data-driven manifestation of LiquidO's topological light ball formation can be best illustrated when comparing the light profiles between opaque and transparent media, as shown in \autoref{fig:lyorbitals}.
    Both light profiles hold common information: scintillator and Cherenkov light emission, the detection system (i.e. fibres) and some common light propagation effects, such as solid-angle.
    A ratio cancels out most of those effects, highlighting the opaque medium differential dispersion photon propagation caused by light scattering.
    The ratio between the light profiles of NoWaSH-Opaque and NoWaSH-Transparent, as a function of the minimal distance between each fibre and the beam spot, exhibits LiquidO's light ball profile caused by its characteristic stochastic light confinement.
    The fractional shape-only distribution of the light ball collection is independent of the light emitted, which is proportional to the energy deposited, as demonstrated with electrons at 0.4, 1.0 and 1.8\,MeV.
    The common profile of the light ball enables us to quantify that 50\% (80\%) [90\%] of the light is detected within a radius of $\sim$2.0\,cm ($\sim$4.0\,cm) [$\sim$5.0\,cm].
    As the shaded regions indicate, less light is collected far from the beam spot, with negligible detection beyond $\sim$10.0\,cm.
    An analytical model, detailed in Appendix~\ref{annexe:analyticalmodel}, fits the data as an illustration guideline of the expected behaviour.
    Despite the model's simple assumptions, including the experimental uncertainties (cyan-band), it provides an excellent description of the data for up to \textgreater90\% of the integrated measured light.
    The remaining 10\% of the light at distances $\geq$5.0\,cm from the beam spot exhibit some discrepancies, likely linked to the over-simplistic assumptions of the model (light ball 3D-symmetry, one scattering length, etc.).
    }
    \label{fig:ratio}
\end{figure*}

\begin{figure*}[!ht]
    \centering
    \includegraphics[width=0.75\linewidth]{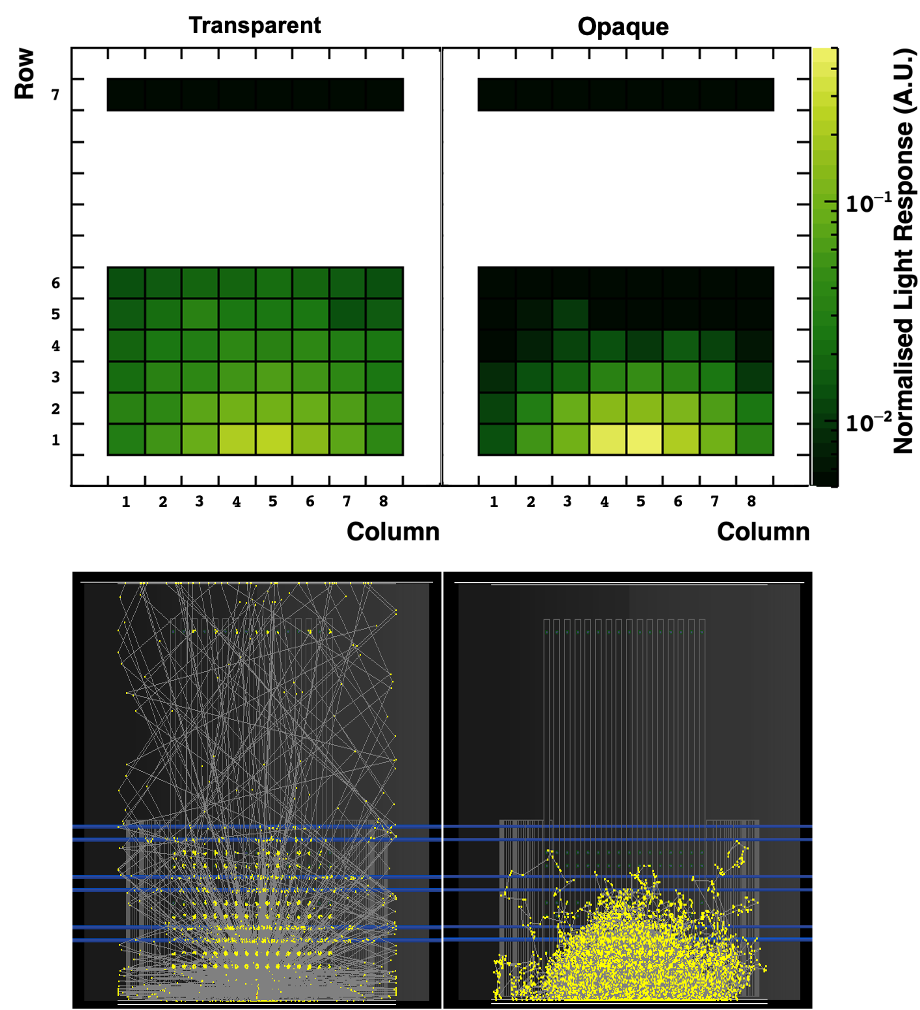}
    \caption{\small 
    {\bf The Mini-LiquidO Detector Hit Pattern.} 
    The upper plots show the average collected light per fibre (z-axis in common log-scale) across the detector for the NoWaSH-Transparent (left) and NoWaSH-Opaque (right) cases.
    The LN is subtracted in both data sets, and the relative uniformity efficiency correction has been applied. 
    Due to the light ball formation in the opaque case, most of the light is detected within the first few fibres.
    These profiles are well reproduced by a customised
    full Geant4-based simulation, detailed in Appendix~\ref{annexe:simulation}, as illustrated in the lower two panels.
    The blue lines (points) represent the fibres running parallel (orthogonal) to the plane of the image. 
    The propagation of photons is graphically indicated where rays (grey lines) are different from photon interactions (yellow dots) due to scattering or absorption.
    Photons mainly interact with the fibres (i.e. detection) and detector walls when transparent (left), while scattering (millimetre scale) dominates when opaque (right).
    See details on simulation configuration in Appendix~\ref{annexe:simulation}.
    Except for the bottom Mylar surface, the stochastic confinement prevents light from hitting the detector walls in the opaque case, which is consistent with the data.
    This data-to-simulation comparison enables an empiric validation of the scattering model in the opaque medium and other expected features, such as the non-symmetrical shape of the light ball due to the bottom surface.
    }
    \label{fig:lightballimage}
\end{figure*}

\subsection{Demonstration of the Stochastic Light Confinement}\label{subsec:confinement}


The main goal of this prototype is to study if and how light is stochastically confined near its creation point (i.e. the beam spot) due to the medium's opacity.
Every fibre samples the light distribution at different distances from the beam spot, and a minimum is obtained by measuring in the direction perpendicular to the fibres. 

The left panel of \autoref{fig:lyorbitals} shows the amount of light detected for 1.8\,MeV electrons as a function of this minimal distance for different media and temperature configurations. 
The amount of light detected per channel is the sum of the total number of PEs observed after LN subtraction and divided by the total number of events. 
All curves are normalised by the same factor, obtained by requiring the light profile for NoWaSH-Opaque (NW 5\textcelsius) to reach unity at the smallest minimum radial distance to the beam spot of 1\,cm.
The fibres with a similar minimal distance to the beam spot (within a few millimetres) are clustered into a single point by averaging their light yields and minimal distances, reducing the 56 readout channels to the 15 points in the graphs.
The NoWaSH-Transparent (NW 40\degreecelsius) curve is consistently lower than the LAB+PPO curve since the former contains 20\% of paraffin wax, which is non-scintillating. 
The right panel shows a subset of the same curves but after normalising their areas to unity, hence the label ``shape-only".
Notably, the NoWaSH-Transparent (NW 40\degreecelsius) and the LAB+PPO curves are almost indistinguishable since light is emitted isotropically.
LAB alone (not shown) exhibits an identical pattern.

A relative uniformity inefficiency correction, detailed in Appendix~\ref{annexe:efficiencycorrection}, is applied to all curves of \autoref{fig:lyorbitals}. 
This correction accounts for correlated features exhibited across all curves for different media configurations, as shown in \autoref{fig:correction}. 
These originate from channel-wise differences in detection efficiency.
The relative uniformity inter-channel readout calibration is performed for each channel by measuring the deviation between the measured light and the expectation with the transparent LAB+PPO data. 
The transparent data can be used for this purpose since the fibre-to-fibre relative response prediction model is dominated by solid angle, as detailed in Appendix~\ref{annexe:efficiencycorrection}.


\autoref{fig:ratio} shows the measured ratio between the NoWaSH-Opaque (NW 5\degreecelsius) and NoWaSH-Transparent (NW 40\degreecelsius) curves from the right side of \autoref{fig:lyorbitals}.
As expected, this ratio is independent of beam energy. 
It is also independent of the relative uniformity correction, which cancels in the ratio.
Some remaining second-order small features may arise due to the binning or other sub-dominant effects.
The simple analytical model of light confinement, detailed in Appendix~\ref{annexe:analyticalmodel}, describes the data well upon fitting, suggesting the simplicity behind LiquidO's ultimate physics model.
This same fact is further corroborated with full simulation relying on a similar model, as detailed in Appendix~\ref{annexe:simulation}.

These results unequivocally demonstrate LiquidO's light ball formation resulting from the light's stochastic confinement around the beam spot. 
The NoWaSH-Opaque sample's large scattering results in a significantly different light collection profile whereby increasingly more light is detected in the fibres closer to the beam spot than the farthest ones.
Specifically, the data show that approximately 65\% more light is detected in the fibres closer to the beam spot with the NoWaSH-Opaque sample, with 50\% (80\%) [90\%] of the light detected within 2\,cm (4\,cm) [5\,cm].
This remarkable increase implies that the light is not lost and rules out an absorption-only scenario.

\begin{table*}[t]
    \centering
    \begin{tabular}{ |l|l|c|c||c| }
\hline
\multirow{3}{*}{}
    & \multirow{2}{*}{Detector Configuration} 
    & \multirow{2}{*}{Scaling Factor}
    & NoWaSH-Opaque
    & NoWaSH-Transparent \\
& 
    &  
    & (PE/MeV) 
    & (PE/MeV) \\ 
\hline
Measured
    & Nominal Response 
    & --
    & \phantom{000}8.4$\,(\pm$10\%)
    & \phantom{000}7.1$\,(\pm$10\%)\\
\hline
\multirow{5}{*}{Scaling} 
    & (1)\phantom{0}$\oplus$~Mylar absorption (loss)
    & 1.67$\times$~or~1.16$\times$ 
    & \phantom{0}14.0
    & \phantom{00}8.2\\

& (2)\phantom{0}$\oplus$~non-scintillator dope (loss)
    & 1.3$\times$ 
    & \phantom{0}18.2
    & \phantom{0}10.7 \\

& (3)\phantom{0}$\oplus$~partial readout (loss)
    & 8.0$\times$ 
    & \phantom{} 145.9
    & \phantom{0}(shadowing) \\

Effects & (4)\phantom{0}$\oplus$~fibre's relative efficiency (loss)
    & [1.5\,,\,2.3]$\times$ 
    & \phantom{0}[218.8\,,\,335.5]
    & \phantom{0}--\\ 


& (5)\phantom{0}$\oplus$~SiPM upgrade (optimisation)
    & 1.5$\times$ 
    & \phantom{0}[328.3\,,\,503.3] 
    & \phantom{0}-- \\
    \cline{2-5}


& (6)\phantom{0}$\oplus$~fibre's attenuation (loss)
    & up to 1.25$\times$
    & neglected 
    & \phantom{0}-- \\

& (7)\phantom{0}$\oplus$~scintillator's absorption (loss)
    & up to 1.45$\times$
    & neglected
    & \phantom{0}-- \\
\hline
\end{tabular}
\caption{\small 
    {\bf Measured Light Response and Scaling.}
    The average total light measured in detected photons (PE) per MeV, or response, is reported here, using the beam energy as normalisation.
    %
    Only the two NoWaSH configurations (opaque and transparent) are considered, as they are directly comparable.
    Each row reports the expected scaling of the measured response due to several detector optimisations.
    The response quoted in each scenario is cumulative, with each entry incorporating the contributions from all previous ones.
    To maximise the scaling accuracy, the system was limited to low responses per pulse ($\leq$10\,PE) so that effects such as pile-up, non-linearities, etc., were mitigated.
    Response systematic uncertainties ($\leq$10\,\%) dominate, as described in the text.
    The transparent setup exhibits a lower relative response ($\sim$85\,\%) due to light escaping the instrumented region or hitting the detector's walls.
    This difference would be even larger were it not for the opaque configuration suffering from a larger fraction of the light absorbed in the Mylar due to multiple reflections, as shown in ~\autoref{fig:lightballimage}.
    The Mylar's absorption is only relevant for LiquidO's surface events (irrelevant in bulk events), which can be simulated and quantified (1).
    %
    Other response losses considered are the non-scintillating fraction (doping) in NoWaSH (2), 
    the partial instrumented readout (3), and
    two scenarios for the fibre's relative efficiency (4).
    A possible upgrade in the light detection system (5) is also illustrated.
    The scaling impact of the fibre's attenuation length (6) and the scintillator's absorption (7) are neglected (see text).
    Many of these issues are due to prototype constraints (low cost, small size, simplicity, impractical calibration, etc.); hence, the scaling enables quantitative response prediction for other LiquidO detector setups.
    In the transparent case, an accurate estimate of the inter-fibre shadowing is impractical, thus compromising the scaling.
    Last, we experimentally prove that LiquidO's opacity leads to both unique topological information and the enhancement of the light collection.
    The specifics are configuration-dependent and may be further optimised.
    The scaled data-driven response shows the potential for a LiquidO-based response from 
    200 up to 
    500\,PE/MeV (optimised), depending on the detection configuration up to order 1\,ton small detector setups.
    The tuned simulation reproduces the data, in agreement with earlier estimates~\cite{LiquidO_2019}.
    Larger LiquidO detectors are possible thanks to the long attenuation length of today's fibre technology.
    }
    \label{tab:abslight}
\end{table*}

The light-ball formation can also be seen in the top row of \autoref{fig:lightballimage}, where the average amount of collected light per channel is shown for the transparent (left) and opaque (right) cases. 
In the latter case, it can be clearly seen that more light is detected closer to the beam spot at the bottom.
It should be noted that the exhibited light-ball shape is a specific feature of this particular configuration, i.e. the chosen media (scattering and absorption) and readout (number of fibres).

\subsection{Absolute Light Response Analysis} \label{subsec:lightamount}

Our prototype was designed to provide accurate information about the amount of light collected in a LiquidO detector not only in terms relative to its transparent counterpart but also in absolute terms. 

The detector's light response, defined as the average integral amount of light observed or expected per 1\,MeV of deposited energy for different media, is summarised in \autoref{tab:abslight}.
The observations in Mini-LiquidO after statistical LN subtraction but before any corrections, including the uniformity efficiency of Appendix~\ref{annexe:efficiencycorrection}, are reported as the nominal configuration.
Only high statistics data collected at 1.8\,MeV are used for these measurements.
Consistent results are obtained with other energies, but the intrinsic non-linearities associated with scintillator quenching~\cite{Birks_1964} and Cherenkov production make their scaling more complex.
The statistical uncertainty is negligible ($\leq$1\,\%), and the observed responses are quoted only with the dominant systematic uncertainty of 10\%.

By design, the response of an optimised LiquidO detector can be predicted by scaling the observations made with this prototype. 
In the opaque case, the scaling can be done with minor biases thanks to the small size of the light ball relative to the dimensions of the detector.
In contrast, the transparent samples suffer from light leakage outside the instrumented volume and losses when light impinges on the detector's walls. 
The reflectivity of Mini-LiquidO's walls was not optimised, and the complex radiator geometry (located $\geq$8\,cm away from the beam spot) causes the response to be lower for transparent media. 

We consider five leading scaling effects in~\autoref{tab:abslight} whose impact has been quantitatively estimated to appreciate the response that could be observed in other LiquidO detectors.
None of these scaling effects are applied in the other results presented in this article. 

\begingroup
    \addtolength{\leftmargini}{-0.5cm}
    \begin{description}

    \item[(1) Mylar Absorption.] 
    The impact of the aluminised Mylar is critical due to its unavoidable very close location to the beam spot.
    Light undergoes multiple reflections on that surface, leading to significant losses. 
    However, its impact varies significantly depending on whether the medium is transparent or opaque.

    In the transparent case, a sizeable fraction of the light hits the Mylar, resulting in a $-$14\% loss estimated from simulation in Mini-LiquidO.
    This number would reduce to $\sim$4\% for a much larger detector.
    With an opaque medium, even more photons hit the Mylar, in some cases multiple times, as illustrated in~\autoref{fig:lightballimage} and \autoref{fig:scattering}.
    Hence, the average loss is found to be ($-$40$\pm$15)\% from simulation, where the uncertainty is dominated by the uncertainty in the true scattering length.
    Consequently, we expect $\sim$1.67$\times$ more light for bulk events in larger LiquidO detectors. 
    However, complex surface effects are important near the detector boundaries, for which these data provide key insights.

    \item[(2) Non-Scintillating Doping.]
    The NoWaSH-Transparent composition has a lower light yield than LAB+PPO, as it contains non-scintillating wax.
    A $20.3\%$ decrease is measured and shown in \autoref{fig:lyorbitals}.
    Since both samples are operated at different temperatures, the colder sample is expected to provide more light.
    In a similar configuration, the light yield was measured to change by -0.4\% per \textcelsius~\cite{Xia_2014}, hence reducing this correction to $\sim$12.3\%.
    Another effect reducing the light yield of the scintillator is caused by quenching due to oxygen contamination~\cite{Birks_1964}, estimated to have a 10\% impact.
    An optimised setup is foreseen to ensure oxygen-stripping and minimise the wax fraction to $\leq$2\%~\cite{yeh_2024_12745269}, yielding virtually no response losses.
    The combined impact of both effects would increase the measured response by $\sim$1.3$\times$.
         
    \item[(3) Partial Readout.] 
    The cost-saving partial instrumentation of the detector readout mentioned in Section~\ref{subsec:setup} leads to a factor of $\sim$8$\times$ less light since only 1 of 8 fibre-ends is instrumented.
    Even if non-instrumented, all fibres actively collect light, shaping the reported light profiles and the light ball formation.
    In the case of the opaque setup, the impact of the shadowing between fibres is expected to be negligible.
    This was validated with simulation and experimentally demonstrated by swapping the fibres.
    However, shadowing is important in the transparent case, and its impact is not reliably quantifiable with simulation due to the complexity of the optical model when dealing with a large number of fibre interfaces and surfaces whose properties are only approximately known; hence, it is not reported.


    \item[(4) Fibre Relative Efficiency.] 
    Our data-driven uniformity correction, detailed in Appendix~\ref{annexe:efficiencycorrection}, evidences an overall light loss and a large spread in response across channels.
    The rather anomalous low average relative efficiency of $\sim$44\% in most channels suggests a general performance issue (i.e. even small damage) at the level of the fibres, which is likely due to excessive handling of the system, typical during prototyping. 
    The optical coupling between fibre and SiPM was independently tested and found stable.
    The estimated impact in light level in the absence of this prototyping-related deterioration leads to a factor of 2.3$\pm$0.2 increase in light, which arises by assuming that all the channels can be made to operate with the same efficiency as the most efficient channel in Mini-LiquidO. 
    Given the uncertainties, a more conservative estimate suggests a smaller 1.5$\pm$0.2 increase.
    None of these estimates accounts for the unknown absolute efficiency of the system.
    
    \item[(5) SiPM Upgrade.] 
    The response may be increased using higher detection efficiency SiPM technology.
    Existing solutions reach up to $\sim$60\%~\cite{hamamatsu-14-16}, while those used in this setup hover around $\sim$40\%.
    This implies a potential $\sim$1.5$\times$ increase in response.

    Admittedly, one of the caveats of using higher efficiency SiPMs today is the increased dark noise.
    While our waveform analysis, detailed in Appendix~\ref{annexe:recozor}, enables excellent dark noise subtraction, this may still be an issue if an extreme energy resolution is needed. 
    Dark noise may also be challenging when self-triggering the detector, which is not true in our setup. 
    Here, larger detectors are expected to require more involved trigger logic. 
    
\end{description}
\endgroup

The data-driven responses resulting from applying the different scaling scenarios to the observations in Mini-LiquidO are summarised in the two right columns of \autoref{tab:abslight}, where they are applied cumulatively. 
These estimates remain conservative since both absorption in the scintillator and attenuation length along the fibres are neglected. 
While absorption has a sizeable impact since the scintillators used were not purified, as discussed in \autoref{tab:liquids}, its accurate quantification suffers from possible additional prototyping limitations.
Instead, light attenuation along the fibres is known and leads to a $\leq$20\% loss (uncorrected) in Mini-LiquidO due to its short fibres (0.5\,m).
The measured response is hence representative of small LiquidO detectors, while the potential for larger detectors is addressed below.

Lastly, other less-known or configuration-dependent elements could change the measured absolute response, such as a possible improved optocoupler efficiency or spectral optimisation of the scintillator-fibre combination. 
Our data does not provide an absolute normalisation to assess their impact.


In summary, the data-driven scaling analysis demonstrates that reasonably high light levels can be obtained in LiquidO detectors at the level of hundreds of PE/MeV, consistent with previous estimations~\cite{LiquidO_2019} and the latest simulation-based analysis, detailed in Appendix~\ref{annexe:simulation}, to reconstruct the scintillator's expected light yield (i.e. photons per MeV) within uncertainties.
For all known effects considered in~\autoref{tab:abslight}, the response may reach up to order $\geq$400\,PE/MeV, depending on the ambitions and cost of the configuration with today's technology, i.e.~minimal R\&D.
This prediction should be valid for small setups ($\leq$1\,ton), although the excellent attenuation length of the B-3 fibres (\textgreater4\,m) grants the potential even in multi-ton setups.
Scaling with the B-3 attenuation length, our data suggests that neutrino detectors with dimensions on the order of 10\,m (i.e. multi-kiloton scale) should feasibly yield responses $\geq$200\,PE/MeV.
That level is directly comparable to today's neutrino scintillator detectors~\cite{DayaBay:2015kir,Allemandou_2018,DoubleChooz_2019,suekane2004overview,ELISEI199753} comfortably operating in energies as low as $\sim$0.5\,MeV.

It should be noted that these estimates are limited to today's state-of-the-art technology, where robust and well-understood solutions exist regarding scintillator formulations, fibres and readout. LiquidO is open to new technological solutions by relaxing the transparency requirement, so dedicated R\&D may further increase its response potential.

\begin{figure*}[!ht]
    \centering
    \includegraphics[width=0.495\linewidth]{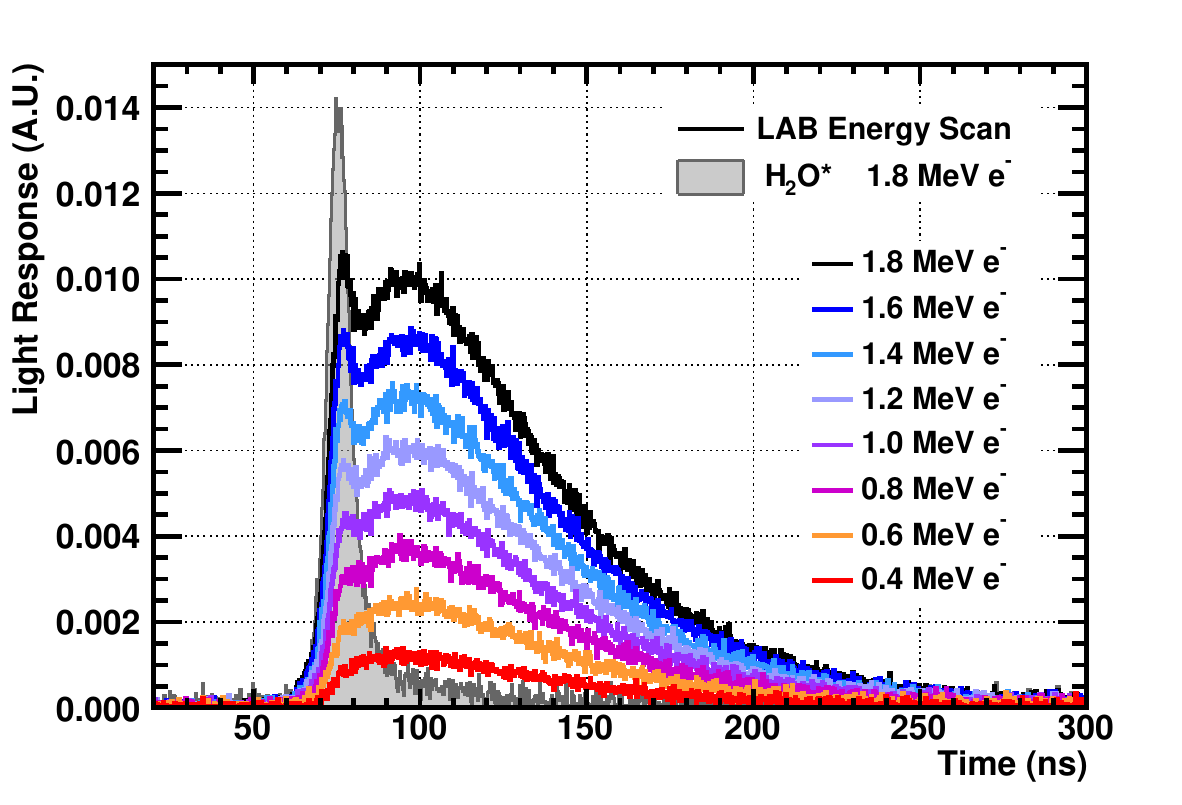}
    \includegraphics[width=0.495\linewidth]{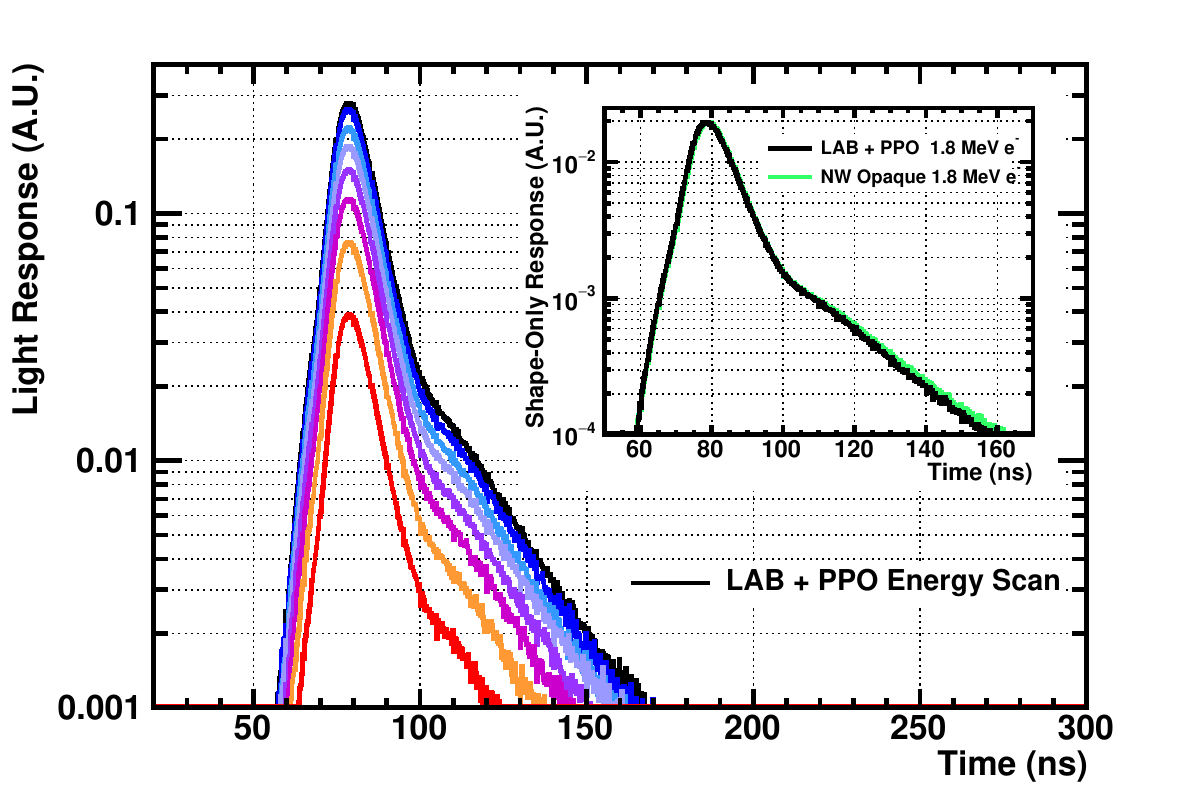}
    \caption{\small 
    {\bf Measured Light Emission Time Distributions.} 
    The two panels show the time distributions of the detected light for different media with different beam energies.
    This illustrates the so-called light emission pulse shapes.
    On the left are the distributions when the detector is operated with pure water or LAB.
    The sharp grey peak corresponds to pure Cherenkov light emitted with 1.8\,MeV electrons in pure water corrected to the LAB's refraction index.
    The coloured curves (energy scan) are obtained with LAB for various electron energies up to 1.8\,MeV. 
    This allows us to visibly distinguish the Cherenkov versus scintillation contributions, thanks to the slow rise of LAB's scintillation.
    The former corresponds to a much smaller peak at 75\,ns.
    The non-linear effect of the $\sim$0.2\,MeV Cherenkov production threshold is also visible. 
    The non-optimised time resolution of Mini-LiquidO is ultimately dominated by fibre deexcitation and light propagation in the fibre.
    No time correction or calibration has been applied.  
    Still, the timing suffices for an excellent statistical separation between both types of light in the LAB sample.
    The light distribution changes dramatically once the nominal amount of PPO (3g/L) is added to the LAB (right).
    The measured time distribution exhibits the expected features: a rapid rise on the order of 1~ns, consistent with the intrinsic fluorescence lifetime of PPO, followed by a decay with two components~\cite{Birks_1964}. These correspond to LAB-to-PPO energy transfer via singlet states ($\sim$70\%) with a characteristic time of $\sim$5~ns and triplet states (~30\%) with a characteristic time of $\sim$25~ns (associated with PPO triplet-triplet annihilation). 
    This demonstrates that LiquidO enables precise characterisation at the nanosecond scale.
    An almost identical pulse shape (green line in inset) is obtained when doping with wax, with a slight deviation consistent with an average time shift per event of $\leq$1 ns, as expected from the opacity.
    The PPO dominates the excitation and deexcitation time constants, and the light output increases to $\sim$50$\times$ ($\sim$6$\times$ only with LAB) relative to the Cherenkov light in water.
    Cherenkov light discrimination becomes impractical in this fast scintillation scenario.
    } 
    \label{fig:timing}
\end{figure*}

\subsection{Timing \& Pulse Shape Analysis}\label{subsec:timing}

One of the most important capabilities required by modern detection techniques is to measure event timing precisely and, if possible, the evolution of energy deposition as a function of time.
While modern electronics allow for sub-nanosecond precision, light detectors are typically limited to a resolution of $\geq$1\,ns, historically dominated by the PMTs' single-PE time resolution. 
A $\sim$1\,ns spread translates into $\sim$20\,cm at the speed of light in transparent media 
irreducibly washing out topological information in energy depositions occurring within a few tens of centimetres.
This is why transparent detectors have found topology impractical unless segmented.

LiquidO changes the paradigm around these limitations in several ways.
First, its capability for self-segmentation unlocks event topology imaging regardless of the time resolution achieved.
Second, the use of SiPMs opens the potential for sub-nanosecond resolutions.
However, the optical readout can include other intrinsic limitations, such as the fibre's time response.
%
Third, while the scattering-dominated medium inherently slows down light propagation, this provides a key advantage: causality-based discrimination between particles moving at the speed of light in a vacuum and photons travelling significantly slower in the medium. 
This feature is a key ingredient of LiquidO's so-called ``{\it energy flow}'' pattern, demonstrated with simulations~\cite{LiquidO_2019} but impractical in this setup due to its small dimensions.

Nonetheless, to accurately characterise the full timing potential of LiquidO, the Mini-LiquidO system was endowed with powerful waveform digitisers, providing up to 320\,ns of history per event, sampled every 312\,ps.
This provides a unique timing test bench for a LiquidO detector exposed to physics events, including possible novel opacity-related features.

Beyond event-wise timing, there is much information within the pulse shape of the light emitted upon medium excitation, and it is pertinent to explore LiquidO's complementary ability to meet the necessary performance for PSD. 
We do this by experimentally characterising the extra smearing arising from the medium opacity to assess the possible deterioration of LiquidO's performance relative to transparent liquid scintillators. 
We also study the ability to quantify the fraction of Cherenkov light in a scintillation pulse. 
This information may enhance PID capability and provide an additional handle to control the detector's calorimetric response. 

These explorations are carried out using the time distributions of electron beam events shown in~\autoref{fig:timing}, which are obtained when the detector is operated with water, LAB and LAB+PPO. 
A measure of LiquidO's intrinsic time resolution can be conservatively obtained from the Cherenkov pulse shape (shaded grey) shown on the left of~\autoref{fig:timing} that is measured in water.
Here, the Cherenkov light excites the B-3 fibres directly, typically integrating its dominant hard-blue, or even ultraviolet, contributions.
No time calibration has been applied.
Hence, possible trigger-jitter effects, inter-channel differences, and the photon time-of-flight from the beam spot to the fibres remain uncorrected.
Some effects amount to sub-ns amplitudes and are difficult to accurately measure experimentally, again, due to the small detector limitations.
Still, from the full width at half-maximum, 
the standard deviation  ($\sigma$) is inferred ($\sim$3.4\,ns) with a sub-dominant tail caused by the fibres.
The ultimate time resolution, if all other effects were corrected, is expected to be limited by the B-3 fibre deexcitation time constant (2.3\,ns) and the propagation along the fibre. 
Faster fibres~\cite{kodama2024performance,kurarayYS1246} can reduce only the time-constant(s) contribution.

The left panel of \autoref{fig:timing} also shows the pulse shape of the LAB solvent during an electron energy scan from 0.4 to 1.8\,MeV (coloured curves).
LAB is not an optimised scintillator, so its light emission is very slow, and its light yield is barely $\sim$6$\times$ that of Cherenkov in water.
To ensure direct comparability in response, the Cherenkov peak was scaled~\cite{Frank_Tamm_1937} considering LAB's refraction index.
The Cherenkov light peak is also clearly distinguishable on the top of the rising and broad ($\sim$100\,ns) LAB pulse shape, which includes very long de-excitation tails.
In addition to the distinctive peak at the same position as in the scaled water data, the unambiguous signature of Cherenkov radiation is demonstrated by its gradual extinction closer to its production threshold at $\sim$0.2\,MeV.
Therefore, Cherenkov light can be detected solely from the detector’s resolution of the early emitted light without exploiting any directional information,
demonstrating LiquidO's excellent time response even before corrections. 

LiquidO's high optical scattering in its opaque medium makes exploitation of the Cherenkov light-cone emission for selection, even for punch-through particles, difficult.
While the time resolution can be enhanced, this is not the main concern in maximising light discrimination.
Another challenge is that a sizeable fraction of Cherenkov light unavoidably excites the scintillator radiatively. 
This is why the LAB exhibits a significantly reduced ($\geq$10$\times$ less) Cherenkov peak for the same 1.8\,MeV excitation.
The B-3 spectral acceptance works like a common filter to enable comparison in both cases.
%



The right panel of~\autoref{fig:timing} shows the impact of the addition of PPO to LAB.
The situation changes significantly: the integral light response increases by roughly 50$\times$ relative to the amount of Cherenkov light in the water-only case, and the PPO dominates the excitation and de-excitation time constants.
The overall pulse shape is significantly faster than LAB alone; hence, the previous Cherenkov light discrimination turns impractical at simple sight.
Active R\&D exists on possible slow rise-time scintillator solutions, such as the LAB alone setup, thus facilitating the distinction of the Cherenkov light even when the light yield is maximised~\cite{Biller_2020}.

Slow scintillators may have detrimental implications for the overall time resolution of detection, which is critical for position reconstruction.
If successful, those solutions are likely to work in LiquidO, as the data shows.

The same pulse shapes are obtained for both LAB+PPO and the NoWaSH-Transparent samples; hence, the presence of wax does not disturb the fluorescence pattern, which is known to be sensitive to the energy transfer of the system.
In the case of NoWaSH-Opaque, shown in the inset, the measured pulse shape is almost identical with a part per thousand statistical shape deviation consistent with an average time-shift per event of $\leq$1\,ns, evidencing the expected impact of scattering; i.e. the opacity.

\section{Conclusion}\label{sec:conclusion}


This publication presents the latest LiquidO results obtained with its largest prototype to date, a 10-litre detector traversed by 208 fibres, in the context of fundamental particle detection.
This prototype combines all elements needed for a complete LiquidO detector, and results are consistent with LiquidO's first proof-of-principle~\cite{LiquidO_2019}. 
The engineering solutions employed remain exploratory, and optimised solutions scalable to larger detectors are under active development.
This experimental configuration benefits from a high-resolution electron beam spectrometer that injects high-purity point-like energy depositions at the bottom of the detector.
The setup successfully demonstrates LiquidO's characteristic feature, namely the stochastic confinement of light around its creation point. 
Specifically, it is found that the light produced by a point-like MeV energy deposition can be efficiently detected ($\geq$90\,\%) within a few centimetres.
The detection scale (i.e. the size of the light ball) can be tuned with the scintillator's scattering mean free path dependence.
In the specific case of the NoWaSH, this is done by varying the temperature. 
The light ball provides the fundamental building block for LiquidO's imaging, granting this technique unprecedented particle-identification capabilities down to the MeV scale. 
The calorimetric properties of the system are also explored, including the production of Cherenkov light at sub-MeV energies due to threshold effects.
The non-optimised time resolution is found to suffice for time-based statistical scintillation to Cherenkov light separation with a performance comparable to other specialised experimental setups.
By design, the direct event-wise light level was kept low (order 10\,PE/MeV) in this prototype, while its scaling was reliably tracked up to hundreds of PE/MeV, a potential performance comparable to that of large, optimised detectors specialised in MeV particle detection.
The potential for even higher light levels is also addressed, which requires further optimisation. 


\section*{Acknowledgements}
{\small
We acknowledge the pivotal support received from the following grants:
i) the ``Chaire Internationale de Recherche Blaise Pascal'' (laureate: Prof.~F.~Suekane) financed by R\'egion \^{I}le-de-France and coordinated by the Fondation de l'\'{E}cole Normale Sup\'erieure (Paris) providing multiple levels of resources for the prototyping of LiquidO;
and 
ii) the Marie Curie Research (grant 707918) fellow M.G. hosted by A.C. at CNRS.
Both grants were hosted at the APC laboratory (Paris).
We would like to strongly thank Hamamatsu Photonics (Japan) and Kuraray corporations, including their representatives in France, for their key support to the LiquidO R\&D activities.

We are thankful to the CNPq/CAPES in Brazil, the McDonald Institute providing FVRF support in Canada, the Charles University in the Czech Republic, the CNRS/IN2P3 in France, the INFN in Italy, the Fundação para a Ciência e a Tecnologia (FCT) in Portugal, the CIEMAT in Spain, the STFC/UKRI/Royal Society in the UK, the University of California at Irvine, Department of Defense, Defense Threat Reduction Agency (HDTRA1-20-2-0002) and the Department of Energy, National Nuclear Security Administration, Consortium for Monitoring, Technology, and Verification (DE-NA0003920), Brookhaven National Laboratory supported by the U.S. Department of Energy under contract DE-AC02-98CH10886 in the USA for their provision of resources.
}

\section*{Authors Contributions}
The Mini-LiquidO prototype detector operations (design and experimental methodology) have been led by the CNRS team, with main contributions (detector system and temperature control, readout, DAQ, and data handling) by the 
APC (Paris),
IJCLab (Orsay),
LP2I (Bordeaux),
and
Subatech (Nantes) laboratories.
The LP2I laboratory provided the MeV single-electron beam,
the MPIK provided the scintillator liquids,
the CIEMAT designed and manufactured the fibre-SiPM connectors,
the team at Sussex University led the simulations (mainly B.C.), 
and the diffusion analytical model was done by JGU Mainz (mainly L.K.).
The LP2I team led the data-taking (mainly M.P.).
The data analysis was led by the IJCLab team (mainly D.N-N.), and cross-checked by the Sussex (mainly W.S.) and CIEMAT(mainly J.A.) teams.
The publication write-up team was constituted by A.C., D.N-N., J.P.O.R., and S.S.
All authors have contributed to this publication through their knowledge and expertise in the LiquidO technology and the final version of the write-up.





\section*{Additional Information}
Correspondence and requests should be addressed to the LiquidO collaboration (\url{LiquidO-Contact-L@in2p3.fr}).


\newpage
\setcounter{section}{0}
\renewcommand*{\thesection}{\Roman{section}}
\renewcommand*{\thesubsection}{\Roman{subsection}}
\section*{APPENDICES}

\subsection{High Purity Event Selection}
\label{annexe:preselection}

The first step in the analysis was removing spurious external triggers not associated with the electron beam.
Those are caused by random coincidences, including readout noise.
External triggering happens by a coincidence of two half-inch Exosens Photonis XP1322 PMTs attached to a 130\,\textmu m thin layer of plastic scintillator~\cite{Ref_BeamSource} situated right below the entrance window of the Mini-LiquidO detector.


All events inconsistent with pure beam energy depositions are rejected.
The rejection criteria of spurious events aim to maximise the sample's high purity at the expense of efficiency since a large amount of statistics was available.
However, this purity is a critical condition to combine and compare data across the data taking (response stability, etc.) and to ensure that the prototype only uses point-like event samples with light originating from the expected beam spot.

Cosmic-ray muons were also removed as part of the aforementioned high-purity selection.
There is a very small probability of muons (low acceptance) happening during our external trigger window since the rate is very low.
The overwhelming majority of these events deposit a much more significant amount of energy in the detector, at least $\sim$2\,MeV per centimetre, compared to electron beam events ($\leq$2\,MeV), as seen by the fibres and the top 3-inch PMT.
This information is used to identify and reject most of them, resulting in negligible contamination of $\leq$0.01\,\% as estimated from data.  

Using a dedicated control run with beam-off, the purity of the samples was estimated to have a small contamination of events on the order of $\sim$1\,\%,  consistent primarily with low-energy electrons.
This background does not compromise the light ball topology analysis but has a small impact on the average light yield quoted, and it is accounted as part of the response systematics to be $\leq$5\,\%.

\subsection{Waveform Reconstruction}
\label{annexe:recozor}

Light pulses recorded by our SiPMs are analysed through a pulse-wise reconstruction and calorimetry method that is based on a detailed understanding of the digitisation process and that evaluates the energy deposited in the detector with better linearity at low energy~\cite{Abe_2014, Hourlier_2016, Pronost_2015}.
This technique enables PE counting and high-precision charge integration, for which extreme baseline control (or zero level) is essential.
The goal is to break down the waveform into its baseline and the corresponding PEs, each characterised by a time and a pulse height.
The algorithm's output consists of a total number of PEs, with the time and pulse height for each, as well as the total charge of the waveform.
The integrated charge allows the reconstruction of a total PE estimator 
if the gain (i.e. the relation between charge per PE) is known, despite possible delicate non-linearity effects~\cite{cabrera2023multicalorimetry,han:tel-03295420}.

\begin{figure*}[!ht]
    \centering
    \includegraphics[width=1.0\linewidth]{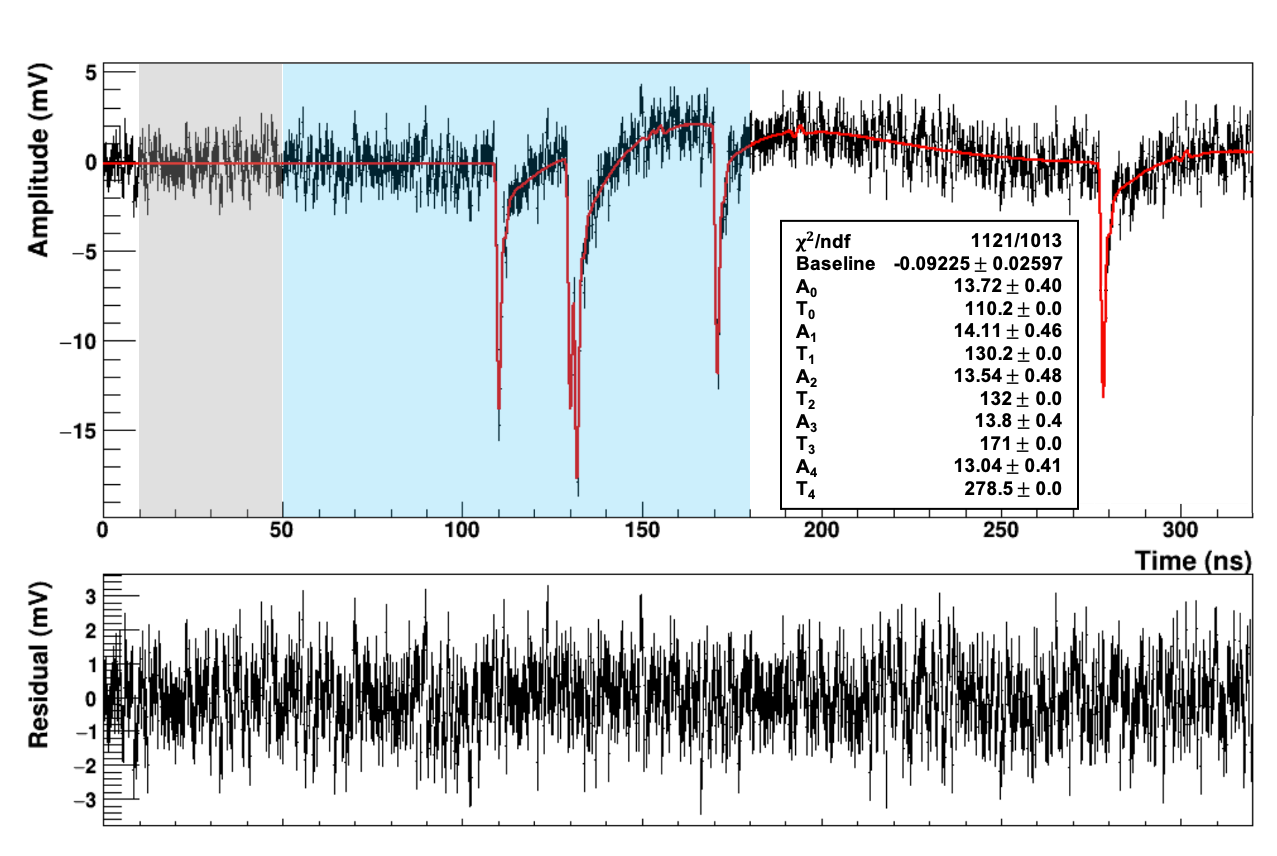}
    \caption{\small 
    \textbf{Pulse Reconstruction Algorithm.}
        An example of the one-channel multi-pulse identification and fitting illustrates the waveform pulse reconstruction algorithm, whose performance depends on the light level.
        The upper plot shows the final outcome of the algorithm, where, in this case, five single PEs have been reconstructed.
        Each PE pulse is fit using a peak-finding algorithm, enabling precise knowledge about each pulse, such as charge (A$_i$; precision $\leq$5\%), time (T$_i$; precision 10\,ps) and the overall common baseline.
        The overshoot maximises the bipolarity of the signals, helping to provide even more accurate baseline information. 
        The fit performance is demonstrated (bottom plot) by the residual between the waveform (raw data) and the fitted outcome function. 
        The grey region indicates the ``pre-time window'' (10-50 ns) used for a coarse baseline estimation and pulse diagnosis. 
        The blue region represents the ``in-time window'' (50 - 180 ns) where most of the signal scintillation light falls.
        The exact position of those windows is relative to the external trigger (electron beam), enabling the statistical combination of events to reach arbitrarily large samples.
        Possible reconstruction non-linearities may occur if the light level increases due to pileup, as illustrated.
        This is one of the reasons why we chose to work at a low-light level per channel regime so that the response scaling estimate, summarised in \autoref{tab:abslight}, suffers from negligible systematic uncertainties.
        In this particular example, the last peak found at 287.5\,ns, beyond the in-time window, is likely to be dark noise from the SiPM, leading to irreducible identical pulses to genuine light, which is dealt with by the statistical subtraction of light noise detailed in Section~\ref{subsec:analysis}.
        }
    \label{fig:recozor}
\end{figure*}

First, a peak-finding method is applied to the waveform to determine how many peaks are present, i.e., to give the minimal PE needed to describe the waveform. 
There can be more than one PE per peak, as shown in~\autoref{fig:recozor}.
Using a data-driven template shape, each peak is treated as a PE.
The template shape accounts inclusively for all relevant experimental effects or features per pulse, such as overshoot and ringing, including its zero amplitude baseline.
This technique enables customisation of the template shape even for each channel, which was done to better accommodate each readout channel's behaviour and pulse shape.
A parametrized shape, or even data, could be part of this template.
The first rough collection of PEs is the basis for the fit function for the waveform.
The final shape is optimised upon fitting to the waveform data, allowing to achieve higher accuracy in amplitude normalisation and position in time.

Critical baseline knowledge is obtained from each pulse, which can be used to diagnose possible anomalous baseline behaviour.
The hardware trigger system is typically delayed by some tens of ns, so a clear baseline sample (no PE from the triggered signal) is obtained at the start of each waveform in the time window referred to as a ``pre-window''. If a random PE (e.g. from dark noise) accidentally falls in this window, the probability of which is not negligible given the high dark-noise rate of order 100,000~Hz, the event is flagged to not bias the baseline determination. 
The observed stability of the baselines generally renders event-by-event determination unnecessary; however, this information is valuable for ensuring the readout's stability and maintaining high data quality standards.  

At this point, the algorithm enters a loop performing a sequence of operations.
First, the residuals between the recorded waveform and the fit function are computed using the baseline.
Second, the time of the most extreme residual is determined, resulting in adding a PE template with the appropriate pulse height to the fit function at that time. 
Third, the new residual is computed, and the loop restarts by incorporating further information (more PE) or tuning the amplitude location of each PE. 
The loop stops when the most extreme residual in the current iteration step is below a given threshold, which is consistent with baseline fluctuations (i.e. no more light pulses).
Finally, the fit function, as given by the algorithm after the end of the loop, is fitted to the waveform. 
The final fit provides the pulse amplitudes and the time of all the peaks found for the waveform, as illustrated in \autoref{fig:recozor}.

Using a data-driven template for the PE shape increases the fit performance significantly compared to relying on a mathematical model. 
The error on PE time after the fit is of the order of tens of picoseconds for a waveform sampling of 312.5\,ps.
Thus, the reconstruction algorithm methodology and waveform sampling are well adapted to extract the maximal energy resolution per PE of the SiPM with negligible additional precision, modulo some systematics effects, typically whenever pileup occurs.

A priori, the resolution per channel is expected to improve stochastically with the number of PE.
However, with more PEs in the same waveform, the complexity of the reconstruction increases and biases may be possible, leading to possible non-stochastic limitations.
Therefore, the ultimate precision of this methodology is unknown, but it is negligible compared to other light-smearing effects within the fibre and scintillator.
In our Mini-LiquidO setup, most of the information relies on counting the average of about one PE per channel and the time of each reconstructed PE.
The light level was kept intentionally low to ensure these potential higher light-yield issues do not distort the reported results, ensuring a robust response determination and scaling as reported in Section~\ref{subsec:lightamount}.

\begin{figure*}[!ht]
    \centering
    \includegraphics[width=1.0\linewidth]{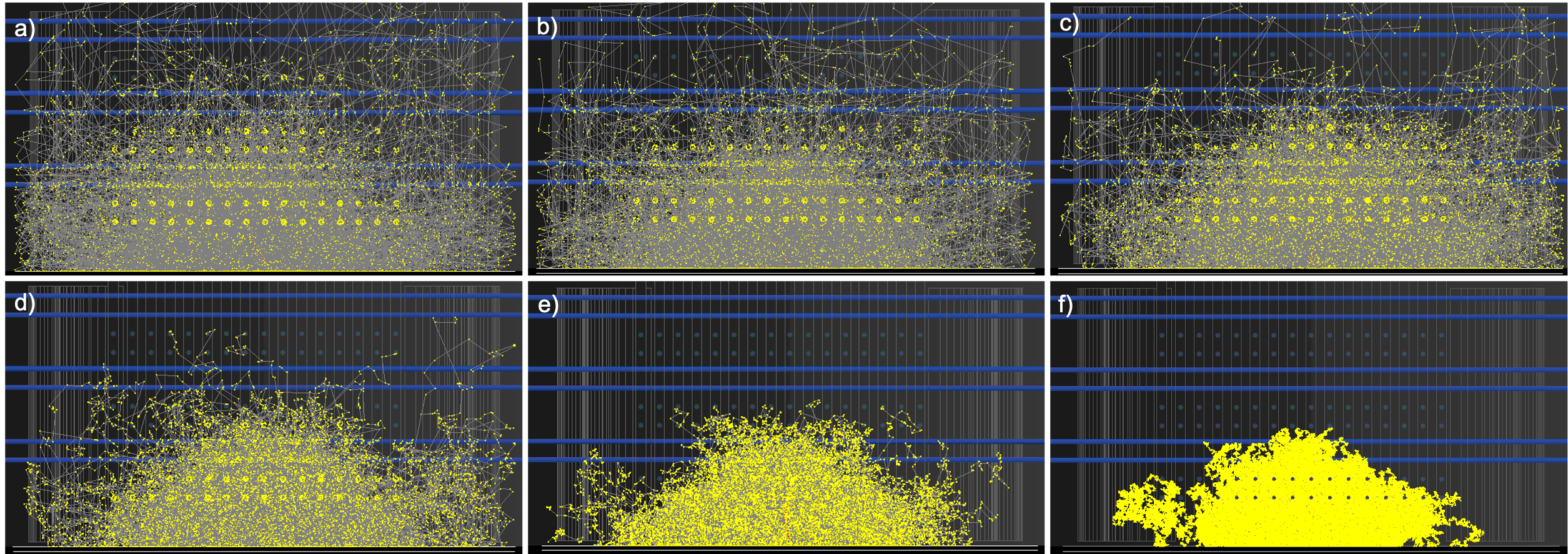}
    \caption{\small 
    {\bf Simulation of LiquidO's Light Ball for Different Scattering Configurations.}
    Geant4 simulation of the detector setup is used to study the impact of the mean free path of elastic light scattering on the light ball formation.
    This empirically estimates the scattering found in the data.
    Several scattering lengths are considered: 
    a) 10.0\,mm, 
    b) 7.5\,mm, 
    c) 5.0\,mm, 
    d) 2.5\,mm, 
    e) 1.0\,mm, 
    and 
    f) 0.1\,mm 
    towards higher opacity medium.
    Due to the simplified model of scattering used in the simulation, a fully meaningful comparison is expected to be impractical.
    The medium is likely to have a distribution of scattering length, leading to an effective superposition of light balls with different dimensions.
    As shown in \autoref{fig:ratio}, the single scattering length approximation is not too far from the data.
    By comparing against data qualitatively, the simulation suggests that the average light scattering length found is close to $\sim$1.0\,mm.
    The simulation tracks (yellow points) each photon interaction in the medium, including the fibres, which can be visually appreciated when the scattering length is larger.
    Given the absorption length in LAB-based scintillators, most of the light is detected with an expected collection efficiencies $\geq$75\%.
    The light ball dimension decreases as the scattering strengthens, keeping all absorption mechanisms constant (fibre, walls and possible losses).
    The optimal scattering length in our setup is order 1.0\,mm, which is $10^{-4}\times$ smaller than traditional transparent configurations.
    The typical rule of thumb for a LiquidO detector is to use a scattering length $\sim$10$\times$ smaller than the fibre pitch if absorption due to losses is minimal.
    }
    \label{fig:scattering}
\end{figure*}

\subsection{Detector's Simulation \& Response Prediction}\label{annexe:simulation}

A full Geant4~\cite{Geant4} based simulation~\cite{Pin_2020} that includes the detailed geometry and composition of the detector was used to provide insight into the principles behind LiquidO detection and to replicate the observed results.
The overall data-to-simulation response agreement was estimated to be $\sim$40\%, dominated by the combined impact of unknowns and uncertainties, as detailed below.

The main goals are to comprehensively understand the average scattering model and predict the measured response with data.
The final response prediction depends to some extent on the scattering model, as explained below.
The unknown scattering length distribution and model specifics are reduced to a single average scattering length in the simulation.
If correct, the simulation should be able to predict the main observables of the system: 
a) the light ball pattern, shown in \autoref{fig:lyorbitals}, \autoref{fig:ratio}, and
\autoref{fig:lightballimage}, together with 
b) the overall measured response, reported in~\autoref{tab:abslight}.
This simplistic approximation in the simulation is also encouraged by our ability to reproduce the overall light-ball pattern even with an analytical model, as described in Appendix~\ref{annexe:analyticalmodel}.
The studies associated with the light ball pattern are summarised in \autoref{fig:scattering}, including a scan in scattering length [0.1,10]\,mm.
Details about the response prediction are summarised below, including a rough breakdown of the leading effects.

The nominal LAB+PPO scintillator is known to yield an order of 10,000 ($\pm$10\%) photons per MeV~\autoref{tab:liquids}.
As described in Section~\ref{subsec:lightamount}, several effects lead to inefficiencies, while some of those are specific to the Mini-LiquidO prototype system, and they were specifically addressed.
Unless otherwise stated, the below numbers refer to \autoref{tab:abslight}.

\begin{figure*}[!ht]
    \centering
    \includegraphics[width=0.49\linewidth]{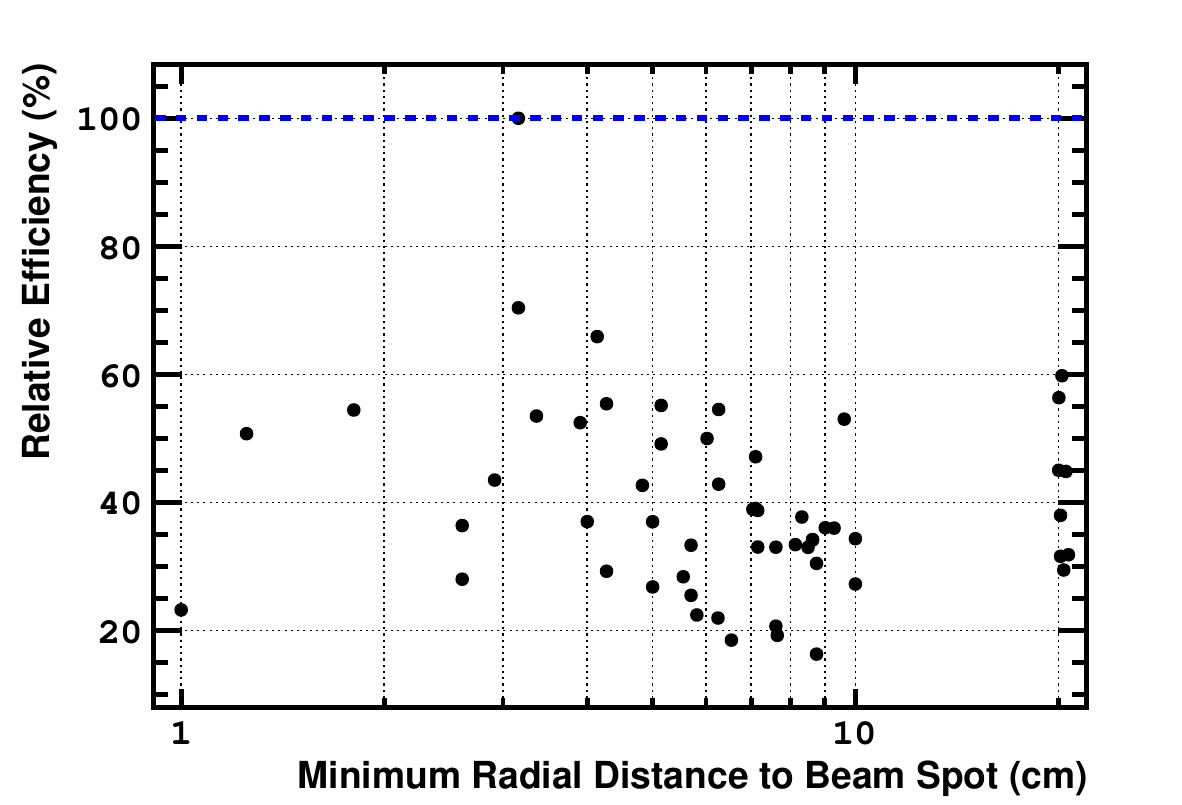}
    \includegraphics[width=0.49\linewidth]{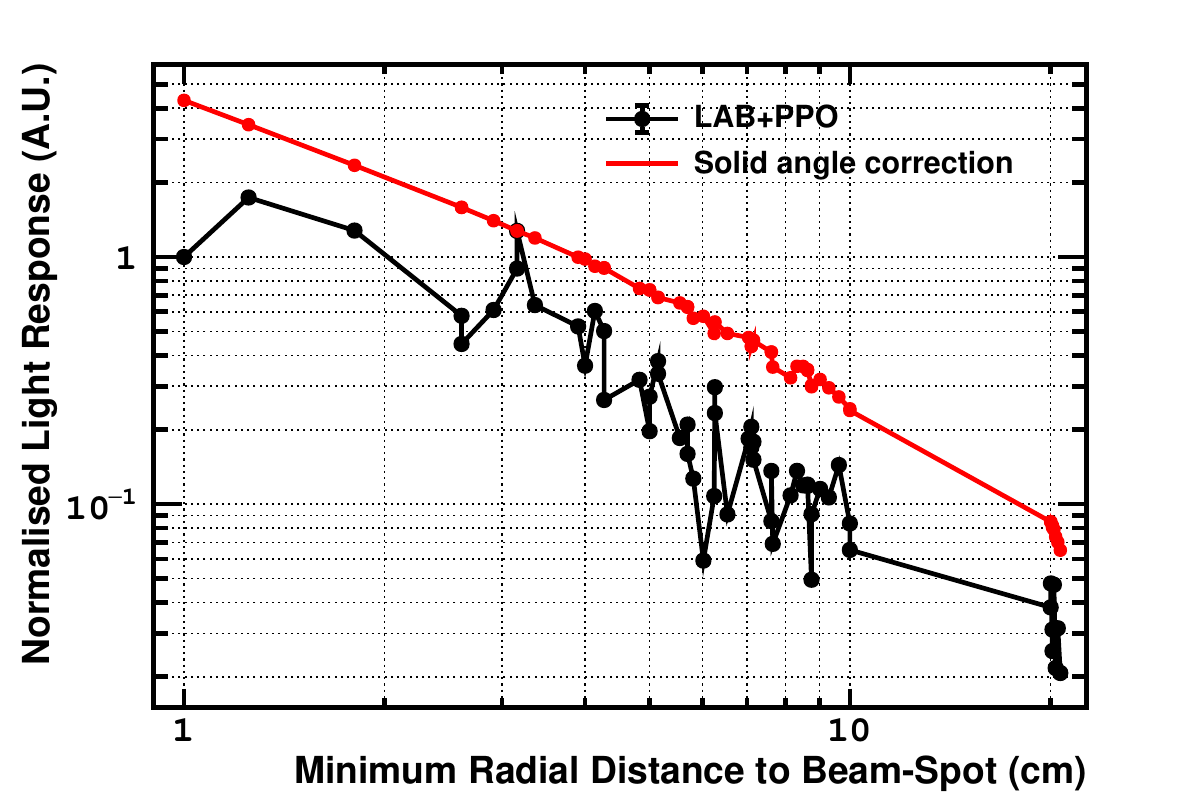}
    \caption{\small 
    {\bf Uniformity Efficiency Correction.}
    The response is expected to exhibit channel-to-channel variations arising from possible fibre-wise or optocoupling-wise variations.
    Due to the simplicity of our setup, the optocoupler scheme was not optimised to yield maximal response or uniformity. 
    Most channels exhibit a large efficiency loss ($\sim$44\%) relative to the channel with the highest response, relatively normalised to 100\%.
    This is assumed to be caused by the significant level of handling of the detector during prototyping, thus causing some degree of degradation.
    Regardless, those variations are typically inclusively corrected by the calibration using a uniform source of response condition.
    Given the limitations of the prototype, this condition was only possible by using transparent control data since photon propagation is simplest, thanks to their isotropic ray-like behaviour leading to solid angle dispersion, as illustrated in \autoref{fig:lyorbitals}.
    We employed the transparent LAB+PPO scintillator as the detection medium, and the impact of the uniformity correction (details in the text) is also illustrated (right).
    The uniformity correction is arbitrarily anchored using the data point with the highest relative efficiency since inefficiencies only manifest as losses, and all fibres are a priori identical from the same production batch.
    Hence, this correction slightly increases the response.
    } 
    \label{fig:correction}
\end{figure*}

The yield quenches 
in the presence of oxygen in the NoWaSH scintillator and its 20\% mass composition of non-scintillating paraffin.
Since the B-3 fibres absorb in the blue, the relevant mean wavelength is $\sim$370\,nm.
An optimised scintillator can have up to 5\,m absorption lengths in this range.
However, since the sample was unfiltered, this is expected to be on the order of $\sim$1\,m, at best.
Using the simulation and considering 1\,mm as the average scattering length, the loss due to absorption is estimated to be -31\%.
This effect was not included as a scaling effect in~\autoref{tab:abslight} due to other possible unknown effects related to prototype handling, which may worsen it.
The scale of this unknown defines the overall inaccuracy uncertainty of the simulation in predicting data.
Exploring the hypothetical case of a purified (filtered) scintillator is interesting.
With an absorption length of \textgreater5\,m, the absorption impact would be as little as $\leq$5\%.
Similarly, the impact of the absorption on the detector walls can be computed.
The loss on the Mylar (average reflectivity 92\%) is dominant.
LiquidO's opacity makes the impact of all other surfaces within a few centimetres of the vicinity (average reflectivity 70\%) relatively negligible thanks to the light being confined close to the beam spot, as illustrated in \autoref{fig:scattering}.

The collection efficiency of LiquidO, as estimated by the simulation, can easily reach \textgreater70\%, depending on the readout configuration.
The overall fibre's light detection efficiency has been estimated to be around $\sim$15\%, including the dominant effect of light trapping in the fibres.
Computing accurately the trapping efficiency is very complex as it depends on the details of the wave-shifting dye absorption length and its concentration. 
The quantitative details of the dye are typically unknown as they are not provided by the manufacturers (i.e. industrial confidential information).
The overall uncertainty associated with the trapping efficiency can be as high as $\sim$30\%. 
Hence, this effect also has an important role in the overall systematic uncertainty of the response prediction.

The impact of the attenuation along the fibres was estimated to be $\sim$80\%, using Kuraray’s data~\cite{Kuraray}.
The detection efficiencies of the SiPM ($\sim$40\%), the reduced 1:8 readout ($\sim$12.5\%) and the measured fibre-coupling system are also considered.
The latter was estimated, using the data-driven uniformity calibration to be $\sim$66\% (conservatively) or $\sim$44\% (normalising to the most efficient fibre), as described in Appendix~\ref{annexe:efficiencycorrection}.

The overall outcome of the simulation predicts an integral response of 10.7$\pm$4.1\,PE per MeV, compared with the data's 8.4$\pm$0.8\,PE per MeV shown in~\autoref{tab:abslight}.
The unknown and, hence, unaccounted absolute inefficiency in data could accommodate the somewhat higher response obtained by the simulation.
However, given the overall uncertainty (dominated by the trapping efficiency and scintillator's absorption in the prototype), the data-to-simulation response agreement is satisfactory.

\subsection{Map of Relative Uniformity Inefficiency}
\label{annexe:efficiencycorrection}

A model that quantifies the variation of light accessible from fibre to fibre was constructed for point-like energy depositions at the bottom of the detector, such as those obtained with low-energy electron beam events.
These variations manifest as a significant non-uniformity in the detected light profile across the detector volume.
The model accounted for relative differences in the light collection efficiencies of all channels, as absolute efficiencies could not be determined without accurately knowing the total amount of light produced in the detector.


The model relies on the fact that the transparent medium has negligible light attenuation relative to the detector size.
Experimentally, this assumption was validated using different transparent media and found to be a good approximation.
The model also neglects reflections along the interior surfaces of the detector.
Accordingly, the only parameter determining the amount of light accessible to each fibre is the solid angle $\Omega$ it covers with respect to the photon source at the bottom. 
This quantity can be accurately estimated by considering the shape of a rectangular plate facing the source, with dimensions equal to the fibre's length and diameter.
The solid angle subtended by such a geometrical object is known and yields
\begin{equation}
    \Omega (a,b,d) = 4 \arctan \left( \frac{\alpha \beta}{\sqrt{1+\alpha ^{2}+\beta ^{2}}} \right), 
\end{equation}
where $(a,b)$ are the dimensions of the rectangular plate, $d$ is its minimal distance from the source, $\alpha = a/(2d)$, and $\beta = b/(2d)$.

The prediction of the relative amount of light seen by all the fibres due to their solid angle coverage whenever the detector is operated with a transparent medium is shown in \autoref{fig:correction} as a function of minimal radial distance to the beam spot, alongside the observations made with LAB+PPO (and validated with LAB data). 
The model's prediction was anchored to the data point associated with the fibre with the highest response, hence resulting in all other points lying below it. 
The deviations from the model were saved as relative response efficiency correction factors, also shown in \autoref{fig:correction}, later used in the analysis as explained in Section~\ref{subsec:confinement}.

The choice of the anchor point for the model can be made arbitrarily as it shifts all the correction factors (correlated), thus having no impact on the light ball shape.
The highest relative response point was chosen as inefficiencies manifest as intrinsic losses in response. Hence, this point holds some physical meaning, illustrating that most fibres had a significantly lower ($\sim$44\,\%) response.
This lower average inefficiency is suspected to be an artefact of prototyping, where constant detection handling and adjustment detrimentally affect the readout quality.
This inefficiency integrates all possible channel-wise differences in coupling (fibres to SiPM) and minor damage per fibre.

\begin{figure*}[!ht]
    \centering
    \includegraphics[width=0.8\linewidth]{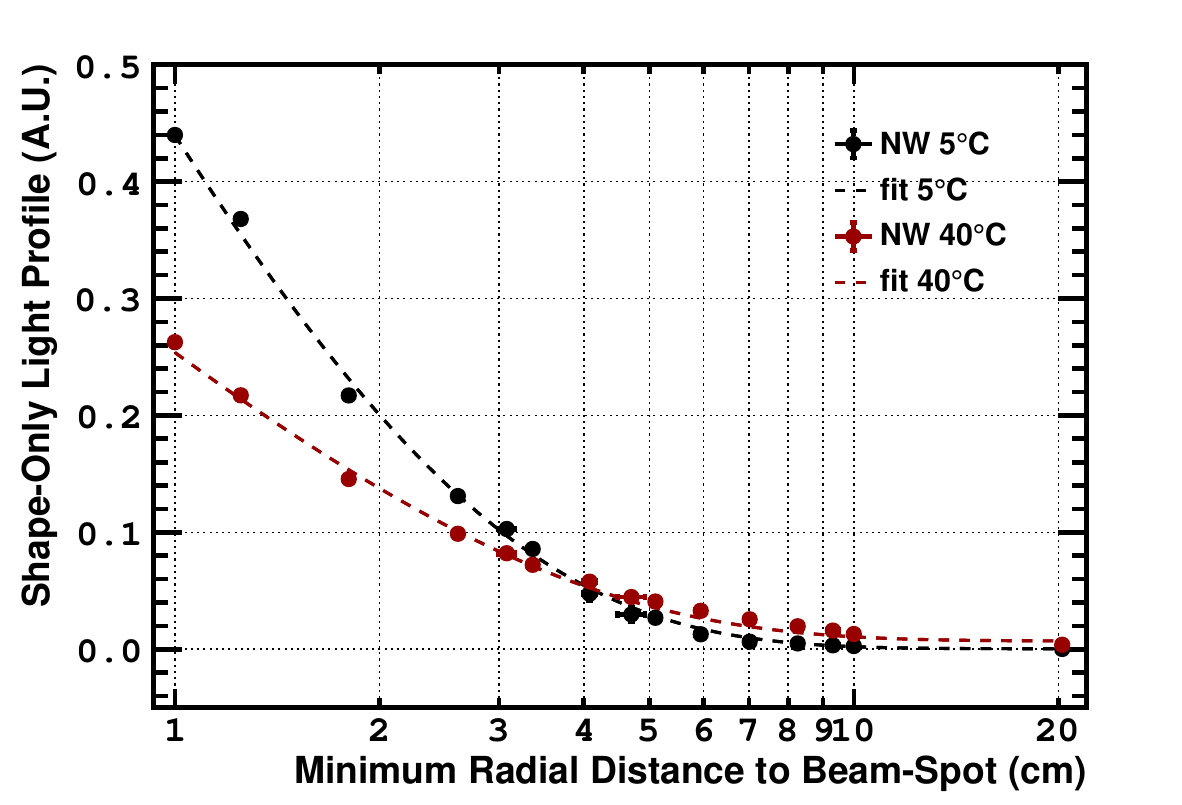}
    \caption{\small 
    {\bf Predicted Profile of Light Ball According to Diffusion-Loss-Model.}
    A simplified diffusion-loss model (details in text) can be used to predict the shape of the light collection and propagation behaviour in a LiquidO detector.
    This is demonstrated in the case of a transparent configuration (brown curve) and the opaque (black curve), once fitted to the corresponding prototype data.
    While this model is not expected to describe the data perfectly, the agreement suggests our understanding of the dominant mechanisms behind LiquidO's light propagation.
    Once fitted, the mean scattering length can be obtained if the light absorption length is known, or vice versa, as the light ball formation is the outcome of a balance between the two.
    In the case of LAB-based scintillation with low intrinsic absorption, the dominant absorption effect is due to the fibres.
    The fitted location of the origin of the light ball is also in excellent agreement, within millimetres, with the expectation. 
    Likewise, the extracted mean free path lengths for absorption and scattering agree with the results derived from the comparisons between simulation and data.
    The ratio of these two curves exhibits the shape of the LiquidO's light ball, as shown in \autoref{fig:ratio} in the context of a shape-only analysis.
    }
    \label{fig:diffusionloss}
\end{figure*}

\subsection{Diffusion-Absorption Analytical Model}\label{annexe:analyticalmodel}


A differential diffusion-absorption (DA) equation can describe light transport through an opaque medium. 
Under the assumption of homogeneity, the equilibrium solutions of this equation for a constant, point-like light source are multivariate Laplace distributions.
Since the probability of being detected in any given detector channel should not depend on the time of a photon's creation, these equilibrium solutions also describe the expected shape of light distributions recorded for transient, point-like light sources in the bulk of opaque detectors. 
Despite its simplicity, we find the analytical model to describe our data well, as shown in \autoref{fig:ratio}, providing additional confirmation that the dominant mechanism behind the observed confinement is due to the stochastic nature of the process.

The DA model assumes that light diffuses due to scattering in an opaque medium until it is absorbed by the optical fibres (i.e. loss through detection) or the medium itself (i.e. genuine loss). 
On length scales much larger than the average scattering lengths of the photons, the light transport in an opaque medium can be described by a differential DA equation:
\begin{equation}
    \dot{n}(\bm{r},t) = a \div \grad n(\bm{r},t) - b\,n(\bm{r},t) + \rho(\bm{r},t) \text{.}
\end{equation}

Here $n(\bm{r})$ is the photon density at position $\bm{r}$ and time $t$, 
$a$ is the diffusion coefficient as required for Fick's law of diffusion~\cite{Fick1855}, 
$b$ is the removal rate, which is assumed to be a constant, 
and $\rho$ is the photon source term. 
In a homogeneous bulk material, $a$ and $b$ are constant in time and space. 
By setting the source term to a delta function in space and the change of photon density to 0, we get the differential equation to determine the equilibrium photon density for a point-like source of photons:
\begin{equation}
    \dot{n}(\bm{r}) = a \div \grad n(\bm{r}) - b n(\bm{r}) + \delta(\bm{r}) \overset{!}{=} 0 \text{.}
\end{equation}

We can ignore the infinitely small extent of the delta function for the bulk of the material and assume that the solution will be rotationally symmetric. 
The equation then simplifies to:
\begin{equation}
    \dot{n}(r) = \qty(a \frac{k}{r} \pdv{r} + a \pdv[2]{r} - b) n(r) \overset{!}{=} 0 \text{,}
\end{equation}
where $k = n_\text{dim} - 1$ is determined by the number of spatial dimensions of the problem, $n_\text{dim} \in \left\{1,2,3\right\}$. 
Multivariate Laplace distributions solve this equation:
\begin{equation}
    n(r) \propto r^\nu K_\nu(\sqrt{2}r/\sigma)\text{,}
\end{equation}
where $K_\nu$ is the modified Bessel function of the second kind, $\sigma = \sqrt{2 a/b}$ is the width of the distribution, and $\nu = (2-n_\text{dim})/2$.
For 1 and 3 spatial dimensions, the $K_\nu$ distribution simplifies somewhat, but in $n_\text{dim}=2$ dimensions the Bessel function remains:
\begin{align}
    n_{1D}(r) &\propto \exp(-\sqrt{2} r/\sigma) \text{,} \\
    n_{2D}(r) &\propto K_0(\sqrt{2} r/\sigma) \text{,}\\
    n_{3D}(r) &\propto \frac{1}{r} \exp(-\sqrt{2} r/\sigma) \text{.}
\end{align}
Note that the projections of multivariate Laplace distributions are again multivariate Laplace distributions of a lower dimension.
So, while fibre-instrumented detectors are very much 3-dimensional bulk structures, the readout of the wavelength-shifting fibres marginalises the distribution along the length of the fibre.
Thus, the most relevant distribution should usually be the 2-dimensional case.

The parameters $a$ and $b$ are macroscopic parameters of the stochastic light diffusion and removal. 
They can be related to the microscopic parameters of the light transport in the medium, though. 
The diffusion coefficient is related to the mean free path length between elastic scatters~\cite{Blum2013}:
\begin{equation}
    a = c \lambda_\text{scatter} / 3\text{,}
\end{equation}
caused by the combined effect of both Rayleigh and Mie scattering, where $c$ is the speed of light. 
Likewise, the absorption rate $b$ is related to the mean free path length before photon absorption:
\begin{equation}
    b = c/\lambda_\text{absorption}\text{.}
\end{equation}
This absorption includes contributions from both the opaque material and the trapping in the wavelength-shifting fibres (expected to be dominant).
The photon transport through the material does not depend on what process exactly absorbs a photon, just the overall rate at which it occurs.
The width of the distribution then relates to the geometric mean of the two mean free path lengths:
\begin{equation}
    \sigma = \sqrt{2\frac{a}{b}} = \sqrt{\frac{2}{3}}\sqrt{\lambda_\text{scatter}\lambda_\text{absorption}}
    \label{eq:diffusionlosswidth}
\end{equation}
This provides a pertinent metric directly applicable to the size of LiquidO's light ball.

In \autoref{fig:diffusionloss}, we compare our data with the analytical diffusion loss model.
From the fit to both NoWaSH transparent and opaque data, we find that our data is best described by a width of the light distribution of $\sigma=28\,\text{mm}$.
As shown in \autoref{fig:scattering}, our simulations show consistent results with our data for a scattering length between 1 and 5\,mm.
Applying \autoref{eq:diffusionlosswidth}, we conclude that the effective mean free path length for combined scintillator and fibre absorption is between 24 and 118\,cm.
This is not the a priori measured absorption of the LAB-based scintillation, which is typically several metres~\cite{Buck_2019}, as the obtained absorption is dominated by the detection of light by fibres (an effective absorption).
This is, however, an interesting characteristic metric of a LiquidO detector, whose light collection efficiency is strongly linked to the value of $\lambda_\text{absorption}$.

Increasing the overall light collection by reducing the absorption in the bulk material will also increase the width of the distribution.
If the width is desired to stay constant, this can be achieved by decreasing the mean free path between scatters, i.e. increasing the concentration of scattering targets.
This perfectly agrees with the results obtained from both simulations, shown in \autoref{fig:diffusionloss}, and data, shown in \autoref{fig:lyorbitals}.

On the other hand, when the light collection is improved by increasing the capture efficiency of the fibres, this will lower the mean free path length before absorption and thus cause a narrower light distribution.
This would then allow the overall density of scattering targets to be lower to keep the overall width constant, which would, in turn, decrease the absorption of light in the bulk material, further boosting the light collection.

\end{multicols*}

\printbibliography 

\end{document}